\documentclass[12pt]{article}
\usepackage[dvips]{color}
\usepackage{amsmath}
\usepackage{graphicx}
\usepackage{subfigure}
\usepackage{epsfig}
\usepackage{amsmath}
\usepackage{cite}
\usepackage{color}
\usepackage{subfigure}
\usepackage{multirow}

\begin{document}
\begin{center}
\Large{\bf Thermodynamic topology of black holes from bulk-boundary, extended, and restricted phase space perspectives}\\
 \small \vspace{0.5cm}
 {\bf Jafar Sadeghi $^{\star}$\footnote {Email:~~~pouriya@ipm.ir}}, \quad
 {\bf Mohmmad  Ali S. Afshar $^{\star}$\footnote {Email:~~~m.a.s.afshar@gmail.com}}\quad
\small \vspace{0.1cm}{\bf Saeed Noori Gashti$^{\dag,\star}$\footnote {Email:~~~saeed.noorigashti@stu.umz.ac.ir}}, \quad\\
 {\bf Mohammad Reza Alipour $^{\star}$\footnote {Email:~~~mr.alipour@stu.umz.ac.ir}}\quad
\\
\vspace{0.1cm}$^{\star}${Department of Physics, Faculty of Basic
Sciences,\\
University of Mazandaran
P. O. Box 47416-95447, Babolsar, Iran}\\
\vspace{0.5cm}$^{\dag}${School of Physics, Damghan University,\\ P. O. Box 3671641167, Damghan, Iran}\\
\small \vspace{0.1cm}
\end{center}
\begin{abstract}
In this article, we investigate the thermodynamic topology of some black holes, namely AdS Reissner Nordstrom (R-N), AdS Einstein-Gauss-Bonnet (EGB), and AdS Einstein-power-Yang-Mills (EPYM), from different frameworks: bulk-boundary (BB) and restricted phase space (RPS). Using the generalized off-shell Helmholtz free energy method, we calculate the thermodynamic topology of the selected black holes in each space separately and determine their topological classifications. We show that the addition of GB terms, dimensions, and other factors do not affect the topological classes of black holes in both spaces. The calculations and plots indicate that the AdS R-N and AdS EGB black holes show similar behavior and their topological numbers sets in both spaces, i.e., BB and RPS, are similar and equal to ($W=+1$). However, AdS EPYM black holes show an interesting behavior. In addition to BBT and RPS, we also consider the extended phase space thermodynamics (EPST) and evaluate the thermodynamic topology for AdS EPYM black hole. The changing ($r-\tau$) in both spaces shows similar behavior. Also, the topological number and the total topological numbers for this black hole in the BB, RPS and EPS thermodynamics are completely same, i,e., $(\omega_{BBT}=\omega_{RPS}=\omega_{EPST}=+1, -1)$ or $W_{BBT}=W_{RPS}=W_{EPST}=0$. An important point is that the Einstein-Yang-Mills black hole has thermodynamic topology equivalence in three spaces. The present result may be due to the non-linear YM charge parameter and the difference between the gauge and gravity corrections in the above black holes
\end{abstract}
\tableofcontents
\section{Introduction}
The study of thermodynamics of black holes is a field of physics that explores the connection between the laws of thermodynamics and the properties of black holes. Black holes have thermodynamic quantities such as entropy and temperature, which are related to classical attributes such as horizon area and surface gravity. Black hole thermodynamics involves integrating general relativity, quantum mechanics and thermodynamics into a comprehensive description of black holes. One of the motivations for studying the thermodynamics of black holes is to understand the statistical mechanics of black holes, which associates entropy with many microstates. This can have a deep impact on the understanding of quantum gravity, leading to the formulation of the holographic principle. Another motivation is to investigate the possible phase transitions and critical phenomena that can occur in black hole systems, such as the Hawking-Page transition between a Schwarzschild black hole and thermal radiation in anti-de Sitter space. There are many interesting and challenging aspects of thermodynamics of black holes, such as the origin and nature of black hole entropy, the information paradox, the quantum corrections to the classical laws, the role of pressure and volume in extended phase space, and the thermodynamics of multi-black hole systems\cite{2,3,a,b,c,d,e,4,5,6,7,8,9,10,11,12,13,14,15,16,17}.\\\\
The relationship between the thermodynamics of black holes and topology is a topic of recent interest in physics. It involves using topological methods and concepts to analyze the phase structure, critical phenomena, and stability of black hole systems. For example, one can use topological defects, winding numbers, and topological charges to classify the local and global properties of the black hole thermodynamic parameter space. One can also use the Brouwer degree, a topological quantity, to reflect the intrinsic properties of the critical points of black hole thermodynamics. Topology can also reveal the similarities and differences between different classes of black holes from the viewpoint of thermodynamics. Some works have been done by using the entropy-temperature (Duan's topological current $\phi$-mapping theory) \cite{1001,1002,1003,1004,1005} and generalized Helmholtz free energy \cite{1006,1007,1008,1009,1010,1011,1012}. Also for further study you can see \cite{1013,1014,1016}.\\\\
Thermodynamics studies in various  phase space including EPS, BB and RPS. A phase space can have different dimensions depending on the number and type of variables that describe the system.
The extended phase space is a formalism that treats the cosmological constant as a thermodynamic variable, related to the pressure by $P=-\Lambda/8\pi G$. This allows for studying the thermodynamic behavior of black holes in anti-de Sitter (AdS) spacetimes, such as critical phenomena, phase transitions and equations of state. The extended phase space also includes the conjugate variable to the pressure, which is the thermodynamic volume V. The extended phase space has the advantage of making the mass of the black hole a homogeneous function of the other variables, satisfying the Euler relation and the Smarr formula.
The RPS is a formalism that excludes the pressure and the volume from the thermodynamic variables, but includes the central charge C of the dual conformal field theory (CFT) and the chemical potential $\mu$. The RPS is motivated by the holographic principle, which states that a gravitational theory in AdS spacetime can be equivalent to a CFT on its boundary. The central charge and the chemical potential are related to the angular momentum and the angular velocity of the black hole by $C = 3J/2 $ and $\mu = \Omega/2 $. The RPS also satisfies the Euler relation and the Smarr formula, but with different coefficients.
The main difference between the extended and RPSs is that they have different sets of thermodynamic variables and different scaling properties. The extended phase space has four pairs of conjugate variables: $(T,S), (P,V), (\Omega,J)$ and $(\phi,Q)$, where T is the temperature, S is the entropy, P is the pressure, V is the volume, $\Omega$ is the angular velocity, J is the angular momentum, $\phi$ is the electric potential and Q is the electric charge. The RPS has three pairs of conjugate variables: (T,S), $(\mu,C)$ and ($\phi$,Q). The extended phase space has a critical point where P, V (the specific volume) and T become equal for all black holes, while the RPS does not have such a point. The extended phase space has a Hawking-Page transition between a thermal AdS state and a large black hole state at a fixed temperature, while the RPS has a Hawking-Page transition between a thermal AdS state and a small black hole state at a fixed chemical potential \cite{103,104,105,106,107,108,109}.\\\\
BB thermodynamics is a formalism that relates the thermodynamic quantities of AdS black holes in the bulk to those of the dual CFT on the boundary via the AdS/CFT correspondence.
This formalism allows one to study the phase transitions and critical points of AdS black holes from both the gravity and the gauge perspectives, and to compare their topological properties. The BB thermodynamic equivalence states that the thermodynamics and phase transitions of AdS black holes in the bulk gravity theory are equivalent to those of the dual CFT on the boundary, from the point of view of topology.
One of the main motivations of BB thermodynamics is to probe some indiscernible phases in the CFT, such as the triple points in the QCD diagram.
Another motivation is to exemplify the gravity-gauge duality in terms of topology, by showing that the bulk and boundary thermodynamics have the same topological numbers for their critical points and phase transitions\cite{111,112}.\\
Finally, our motivation to study this article was because the methods of ”Wei”et al., which use topological thermodynamics to classify black holes, has been applied in several articles, either based on the temperature model \cite{1001,1002,1003,1004,1005,2005,3005,1a,2a,3a,4a} or based on the generalized free energy \cite{1c,1007,1008,1009,1010,1011,1b,2b,3b,1012,1013,1014,1016,101616}. However, we wanted to know how this thermodynamic topology of black holes changes considering different frameworks, such as bulk boundary, extended, and restricted phase space, and also what are the physical implications and significance of the thermodynamic topology of black holes with respect to holography principle and the phase transitions?\\Therefore, we have organized the article in the following form : In section 2, we give a brief explanation of thermodynamic topology from the perspective of generalized Helmholtz free energy. In section 3, we brief review of the AdS R-N black holes, AdS EGB black holes and AdS EPYM black holes. In section 4, we examine the thermodynamic topology for the AdS R-N, AdS EGB and AdS EPYM black holes in the extended phase space and do a similar analysis for the RPS in section 5. We discuss the results in section 6 and finally, we describe the results in detail in section 7.

\section{Topology of black holes thermodynamic}
Thermodynamic topology is a method that introduces topology to the study of black hole thermodynamics by assigning a topological numbers to each zero point in the phase diagram.
The topological number is defined as the residue of the generalized Helmholtz free energy at the critical point, and it can reveal some new features and classifications of black hole thermodynamics that are not captured by conventional methods . For example, it can distinguish between conventional and novel critical points, which have different implications for the first-order phase transition. Helmholtz off-shell free energy is a generalization of the Helmholtz free energy that allows one to consider non-equilibrium states of the system. The Helmholtz free energy is defined as the internal energy of the system minus the product of the temperature times the entropy of the system, and it measures the useful work obtainable from a closed thermodynamic system at a constant temperature and volume. The Helmholtz off-shell free energy is defined as the Legendre transform of the internal energy with respect to the entropy, and it can be written as:
\begin{equation}\label{1}
F(S,V,Y) = U(S,V,Y) - TS
\end{equation}
where $S$ is the entropy, $V$ is the volume, $Y$ are other extensive variables, $T$ is the temperature, and $U$ is the internal energy.
The Helmholtz off-shell free energy can be used to study the phase transitions and critical points of black holes in different contexts, such as AdS or dS space, with or without electric charge, nonlinear electromagnetic fields, etc . We use different quantities to introduce the thermodynamic properties of black holes.Researchers have so far presented different methods for calculating the generalized Helmholtz free energy according to different perspectives which we can refer to some of them.\cite{10061,10062,10063,10064,10065}.
We had to examine the topological structure of thermodynamics based on \cite{1c}. Accordingly, we write the form of the generalized Helmholtz free energy in the following form \cite{1c,1007,1008,1009,1010,1011,1b,2b,1012,1013,1014,1016,101616},
\begin{equation}\label{2}
\mathcal{F}=M-\frac{S}{\tau}.
\end{equation}
Where $\tau$ and T (inverse of $\tau$) are the Euclidean time period and the temperature of the ensemble. The vector $\phi$ is given by,
\begin{equation}\label{3}
\phi=(\phi^{r_h}, \phi^{\Theta})=\big(\frac{\partial\mathcal{F}}{\partial r_{h}},-\cot\Theta\csc\Theta\big).
\end{equation}
The vector $\phi^{\Theta}$ has infinite magnitude and points away from the origin when $\Theta$ is either 0 or $\pi$. The variables $r_{h}$ and $\Theta$ can take any values from 0 to infinity and from 0 to $\pi$, respectively.
also here, we can  rewrite the vector as $\phi=||\phi||e^{i\Theta}$, where $||\phi||=\sqrt{\phi^a\phi^a}$, or $\phi = \phi^{r_h} + i\phi^\Theta$.
 Based on this, the normalized vector is defined as,
 \begin{equation}\label{2}
n^a=\frac{\phi^a}{||\phi||},
\end{equation}
 where $a=1,2$  and  $(\phi^1=\phi^{r_h})$ , $(\phi^2=\phi^\Theta)$.
Now we introduce our antisymmetric superpotential as follows,
 \begin{center}
 $V^{\mu\nu}=\frac{1}{2\pi} \epsilon^{\mu\nu\rho} \epsilon_{ab}n^a\partial_\rho n^b,\hspace{0.3cm}\mu,\nu,\rho=0,1,2$,
\end{center}
 We can use Duan's theory to define a current that depends on the topology of the vector field $\phi$ as follows,
\begin{equation}\label{4}
j^{\mu}=\frac{1}{2\pi}\varepsilon^{\mu\nu\rho}\varepsilon_{ab}\partial_{\nu}n^{a}\partial_{\rho}n^{b},\hspace{1cm}\mu,\nu,\rho=0,1,2
\end{equation}
$n$ is a unit vector that has the same direction as $\phi$. The components of $n$ are the ratios of the components of $\phi$ and its magnitude. According to Noether's theorem, the topological currents do not change and we will have,
\begin{equation}\label{5}
\partial_{\mu}j^{\mu}=0,
\end{equation}
we can write the $j^{\mu}$,
\begin{equation}\label{6}
j^{\mu}=\delta^{2}(\phi) J^{\mu}(\frac{\phi}{x}).
\end{equation}
Also,
\begin{equation}\label{7}
\varepsilon^{ab}J^{\mu}(\frac{\phi}{x})=\varepsilon^{\mu\nu\rho}\partial_{\nu}\phi^{a}\partial_{\rho}\phi^{b}
\end{equation}
The Jacobi vector is a special case of the generalized Jacobi vector when $\mu$ is zero. The determinant of the Jacobi matrix is equal to $J^{0}\big(\frac{\phi}{x}\big)=\frac{\partial(\phi^1,\phi^2)}{\partial(x^1,x^2)}$. Equation (5) shows that $j^{\mu}$ is zero everywhere except at the origin. We can find the topological number or winding number of $W$, which is given by,
\begin{equation}\label{8}
W=\int_{\Sigma}j^{0}d^2 x=\Sigma_{i=1}^{n}\beta_{i}\eta_{i}=\Sigma_{i=1}^{n}\omega_{i}.
\end{equation}
$\beta_i$ is a topological invariant that measures the degree of linking between the vector $\phi^a$ and the origin in the $\phi$ space. The $\phi$ space is a two-dimensional space where the vector $\phi^a$ lives. The $x$ space is a four-dimensional space where the coordinates $x^\mu$ live. The zero point $z_i$ is a point in the $x$ space where the vector $\phi^a$ vanishes. When we look at a small neighborhood of the zero point $z_i$, we can map it to a sphere in the $\phi$ space. The vector $\phi^a$ then traces a curve on the sphere, and we can count how many times this curve winds around the origin. This number is the Hopf index $\beta_i$, and it is always positive. The sign of $\eta_i$ is given by the sign of the topological current $j^0(\phi/x)_{z_i}$, which is a scalar quantity that depends on the vector $\phi^a$ and its derivatives. The topological current is zero everywhere except at the zero points, where it becomes infinite. The quantity $\omega_i$ is another topological invariant that measures the degree of rotation of the vector $\phi^a$ around the zero point $z_i$. To calculate it, we need to choose a closed curve $\Sigma$ in the $x$ space that encloses the zero point $z_i$. We can then map this curve to a circle in the $\phi$ space, and see how many times the vector $\phi^a$ rotates along this circle. This number is the winding number $\omega_i$, and it can be positive or negative. The shape of the curve $\Sigma$ does not affect the value of the winding number, as long as it does not cross any other zero points.

\section{Thermodynamics of Black holes}
Thermodynamics of black holes is a field of physics that explores the connection between the laws of thermodynamics and the properties of black holes. Black holes have thermodynamic quantities such as "entropy" and "temperature", which are related to classical attributes such as "horizon area" and "surface gravity". Black hole thermodynamics involves integrating general relativity, quantum mechanics and thermodynamics into a comprehensive description of black holes. The "second law of thermodynamics" requires that black holes have entropy. If black holes carried no entropy, it would be possible to violate the second law by throwing mass into the black hole. The increase of the entropy of the black hole more than compensates for the decrease of the entropy carried by the object that was swallowed. The "Bekenstein–Hawking formula" gives the entropy of a black hole as proportional to its horizon area: $S_{BH} = \frac{k_B c^3 A}{4 G \hbar}$ where $k_B$, $c$, $A$ ,$G$ and $\hbar$ determine the Boltzmann constant, speed of light, area, gravitational constant, and  reduced Planck constant, respectibely. Black holes emit "thermal Hawking radiation" corresponding to a certain temperature (Hawking temperature), which is inversely proportional to their mass: $T_H = \frac{\hbar c^3}{8 \pi G M k_B}$ where $M$ is the mass of the black hole. There are more complex and interesting multi-black hole systems, such as those involving cosmic strings or acceleration, which can also be studied using thermodynamic principles.

\subsection{Case I: Thermodynamics of AdS Reissner-Nordstrom black holes}
AdS R-N black holes are black holes in anti-de Sitter space with an Abelian electric charge. They are solutions of the Einstein-Maxwell equations with a negative cosmological constant. They have two horizons, an event horizon and a Cauchy horizon, which depend on the mass and charge of the black hole. They have a Hawking temperature, an entropy, and an electrical potential, which satisfy the first law of thermodynamics. They also exhibit a phase transition from a small black hole to a large black hole, or vice versa, when the temperature or pressure reaches a critical value. This phase transition is similar to the liquid-gas phase transition of a Van der Waals fluid. Thermodynamics of AdS R-N black holes is a topic that studies the thermodynamic properties and phase transitions of charged black holes in anti-de Sitter space. The "metric" of an AdS R-N black hole is given by\cite{113}:
\begin{equation}\label{9}
ds^2 = -f(r) dt^2 + \frac{dr^2}{f(r)} + r^2 d\Omega^2
\end{equation}
where
\begin{equation}\label{10}
f(r) = 1 - \frac{2M}{r} + \frac{q^2}{r^2} + \frac{r^2}{l^2}
\end{equation}
where $M$ is the mass, $q$ is the charge, and $l$ is the AdS radius. This metric describes the spacetime geometry around the black hole, which has a singularity at $r=0$ and one or two horizons depending on the values of $M$, $q$ and $l$. The horizons are the surfaces where $f(r_h)=0$, and the outermost one is called the event horizon, which determines the size of the black hole. The "temperature" of an AdS R-N black hole is given by:
\begin{equation}\label{11}
T = \frac{f'(r_h)}{4\pi} = \frac{1}{4\pi r_h} \left( 3 - \frac{q^2}{r_h^2} + \frac{3r_h^2}{l^2} \right),
\end{equation}
where $r_h$ is the horizon radius. This temperature is related to the surface gravity of the black hole, which measures the strength of the gravitational field at the horizon. It also determines the rate of Hawking radiation emitted by the black hole, which is a quantum effect that causes the black hole to lose mass and charge over time. The "entropy" of an AdS R-N black hole is given by:
\begin{equation}\label{12}
S = \frac{k_B A}{4 G \hbar} = \frac{k_B \pi r_h^2}{G \hbar}
\end{equation}
This entropy is proportional to the horizon area, which follows from the Bekenstein-Hawking formula that applies to any black hole. It also satisfies the second law of thermodynamics, which states that the total entropy of a system never decreases in any physical process.

\subsection{Case II: Thermodynamics of AdS Einstein-Gauss-Bonnet black holes}
AdS EGB black holes are black holes in anti-de Sitter space with a higher-order curvature correction term. They are solutions of the EGB equations with a negative cosmological constant. They have one or two horizons, depending on the values of the mass, charge, and Gauss-Bonnet coupling constant of the black hole. They have a Hawking temperature, an entropy, and an electrical potential, which satisfy the first law of thermodynamics. They also exhibit a phase transition from a small black hole to a large black hole, or vice versa, when the temperature or pressure reaches a critical value. This phase transition is different from the liquid-gas phase transition of a Van der Waals fluid, and depends on the sign and magnitude of the Gauss-Bonnet coupling constant.  The thermodynamics of AdS EGB black holes is concerned with the properties and behavior of these solutions in relation to their mass, temperature, entropy, heat capacity and free energy. The EGB theory in D dimensions is described by the action\cite{1113,114},
\begin{equation}\label{13}
\mathcal{S}=\frac{1}{16\pi}\int d^{D}x\sqrt{-g}|R+\alpha\mathcal{L}|,
\end{equation}
where
\begin{equation}\label{14}
\mathcal{L}=R^2-4R_{\mu\nu}R^{\mu\nu}+R_{\mu\nu\rho\sigma}R^{\mu\nu\rho\sigma},
\end{equation}
We use the notation $R$,  $R_{\mu\nu} $ and $R_{\mu\nu\rho\sigma}$ for the Ricci scalar, the Ricci tensor, and the Riemann tensor, respectively. The metric $g_{\mu\nu}$ has a determinant that we also call $g$.
The Gauss-Bonnet term has no effect on the dynamics in four dimensions ($D=4$), because its integral is a topological invariant. However, we can change the coupling constant by scaling it as,
\begin{equation}\label{15}
\alpha\rightarrow\frac{\alpha}{D-4},
\end{equation}
The metric of an AdS EGB black hole with spherical symmetry can be written as:
\begin{equation}\label{16}
ds^2 = -f(r)dt^2 + \frac{dr^2}{f(r)} + r^2 d\Omega_{D-2}^2
\end{equation}
where $d\Omega_{D-2}^2$ is the metric on a unit $(D-2)$-sphere and $f(r)$ is the metric function for the 4D AdS EGB black hole is given by\cite{114},
$$f(r)=1+\frac{r^2}{2\alpha}\bigg(1-\sqrt{1+4\alpha\bigg(\frac{2M}{r^3}-\frac{q^2 }{r^4}-\frac{1}{l^2}\bigg)}\bigg)$$
Here, $M$ is the mass parameter, $\Lambda$ is the cosmological constant, $\alpha$ is the Gauss-Bonnet coupling constant. The "temperature" of a black hole is given by:
\begin{equation}\label{17}
T =\frac{f'(r_h)}{4 \pi}
\end{equation}

\subsection{Case III: Thermodynamics of AdS Einstein-power-Yang-Mills black holes}
AdS Einstein-power-Yang-Mills black holes are a class of black hole solutions in higher-dimensional gravity theories that involve a non-linear generalization of the Yang-Mills field. The Yang-Mills field is a type of gauge field that describes the interactions of elementary particles in quantum field theory. The power-Yang-Mills field is defined by a Lagrangian density that contains a power of the Yang-Mills field strength tensor. The AdS Einstein-power-Yang-Mills black holes are solutions of the Einstein field equations with a negative cosmological constant and a power-Yang-Mills source term. They have interesting properties such as horizon structure, thermodynamics, phase transitions, and Joule-Thomson expansion. The "metric" of an AdS EPYM black hole is given by\cite{1114,115}:
\begin{equation}\label{18}
ds^2 = -f(r) dt^2 + \frac{dr^2}{f(r)} + r^2 d\Omega_2^2
\end{equation}
where $$f(r)=1-\frac{2M}{r}+\frac{r^{2}}{l^2}+\frac{(2q^2)^\gamma}{2(4\gamma-3)r^{4\gamma-2}}$$
where $M$ is the mass, $q$ is the Yang-Mills charge, and  $\gamma$ is the non-linear YM charge parameter. The black hole has horizons, an event horizon at $r_h$ and a Cauchy horizon at $r_-$, which are determined by the roots of $f({r_h}) = 0$. The "temperature" of an AdS EPYM black hole is given by\cite{1114}:
\begin{equation}\label{19}
T = \frac{1}{4\pi r_h}\bigg(1+8\pi Pr_{h}^{2}-\frac{(2q^{2})^\gamma}{2r_{h}^{(4\gamma-2)}}\bigg)
\end{equation}
where $r_h$ is the event horizon radius. The "entropy" of an AdS EPYM black hole is given by:
\begin{equation}\label{20}
S =\frac{A}{4},
\end{equation}
where $A$ is the horizon area. This entropy describes the measure of disorder or information loss associated with the black hole, which depends only on the horizon area according to the Bekenstein-Hawking formula.

\section{Thermodynamics of Bulk-Boundary}
BB thermodynamics is a formalism that relates the thermodynamic quantities of AdS black holes in the bulk to those of the dual CFT on the boundary via the AdS/CFT correspondence. BB thermodynamic equivalence is a concept that states that the thermodynamics and phase transitions of AdS black holes in the bulk gravity theory are equivalent to those of the dual CFT on the boundary, from the point of view of topology. Topological methods can be used to study the critical points and phase transitions of black hole thermodynamics by assigning a topological number to each critical point in the phase diagram. A topological number is a number that characterizes the type and order of the phase transition at a critical point. It can be calculated by using the residue method or Duan's topological current-mapping theory. In this section, we aim to explore the thermodynamic topology of the three black holes mentioned above in the BB framework. We explain the details of the calculations in the following subsections
\subsection{Case I}
With respect to equation (10), we will have,
$$f(r) = 1 - \frac{2GM}{r} + \frac{Gq^2}{r^2} + \frac{r^2}{l^2}$$
The entropy and radius of AdS are calculated in the following form
\begin{equation}\label{21}
\begin{split}
&S  = \frac{r_h^{2} \pi}{G},\\
&l = \frac{\sqrt{6}\, \sqrt{\frac{1}{P G \pi}}}{4}
\end{split}
\end{equation}
The Hawking temperature for AdS R-N black hole is rewritten in the following form according to the equation (11),
\begin{equation}\label{22}
T =\frac{2 \left(-\frac{3 q^{2}}{8 P \pi}+\frac{3 r_h^{2}}{8 P G \pi}+3 r_h^{4}\right) P G}{3 r_h^{3}}
\end{equation}
Since $G$ is considered a variable in this space, so
\begin{equation}\label{23}
G =\frac{r_h^{2}}{8 P \pi  r_h^{4}+3 q^{2}}.
\end{equation}
Also, according to the mentioned equations, mass and Helmholtz free energy for this black hole are calculated in the following form
\begin{equation}\label{24}
M =\frac{4 \left(\frac{3 q^{2}}{8 P \pi}+\frac{3 r_h^{2}}{8 P G \pi}+r_h^{4}\right) P \pi}{3 r_h}
\end{equation}
and
\begin{equation}\label{25}
\mathcal{F} =\frac{4 \left(\frac{3 q^{2}}{8 P \pi}+\frac{3 r_h^{2}}{8 P G \pi}+r_h^{4}\right) P \pi}{3 r_h}-\frac{r_h^{2} \pi}{G \tau}
\end{equation}
Two vector field $\phi^{r_h}$ and $\phi^{\Theta}$ with respect to equation (3) and mentioned concepts are calculated as,
\begin{equation}\label{26}
\begin{split}
&\phi^{r_h} =\frac{\left(8 P G \pi  r_h^{4}-q^{2} G +r_h^{2}\right) \tau -4 \pi  r_h^{3}}{2 G \,r_h^{2} \tau},\\
&\phi^{\Theta} = -\frac{\cos \! \left(\Theta \right)}{\sin \! \left(\Theta \right)^{2}}
\end{split}
\end{equation}
Now, we can calculate considering the equation (26) and $\phi^{r_h} =\frac{\partial\mathcal{F}}{\partial {r_h}}$
\begin{equation}\label{27}
\tau =\frac{4 \pi  r_h^{3}}{8 P G \pi  r_h^{4}-q^{2} G +r_h^{2}}
\end{equation}
According to the above concepts, we can express the changes of ($\theta$) and ($\tau$) in terms of r. Later, we will describe the results in detail
\begin{figure}[h!]
 \begin{center}
 \subfigure[]{
 \includegraphics[height=5.5cm,width=6cm]{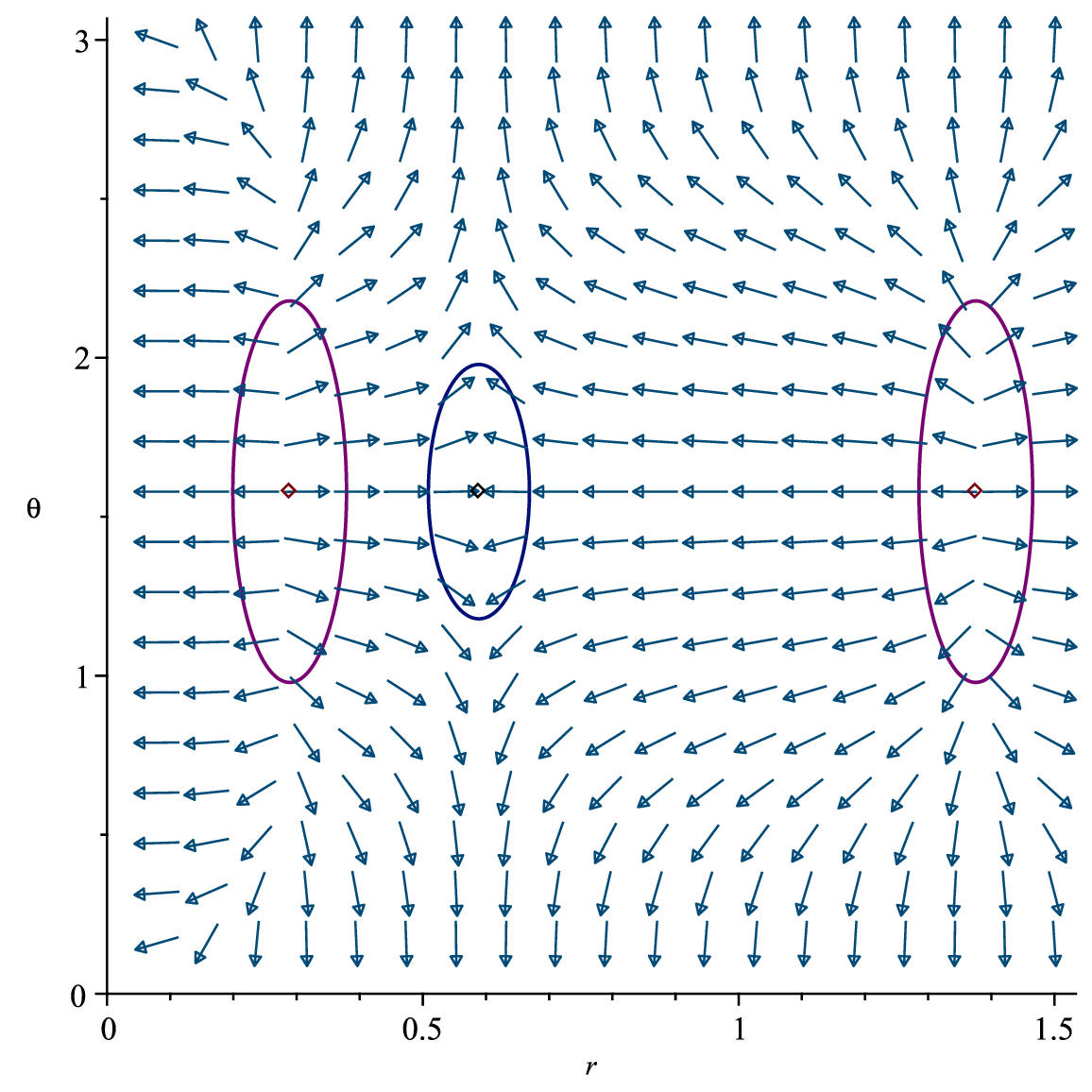}
 \label{1a}}
 \subfigure[]{
 \includegraphics[height=5.5cm,width=6cm]{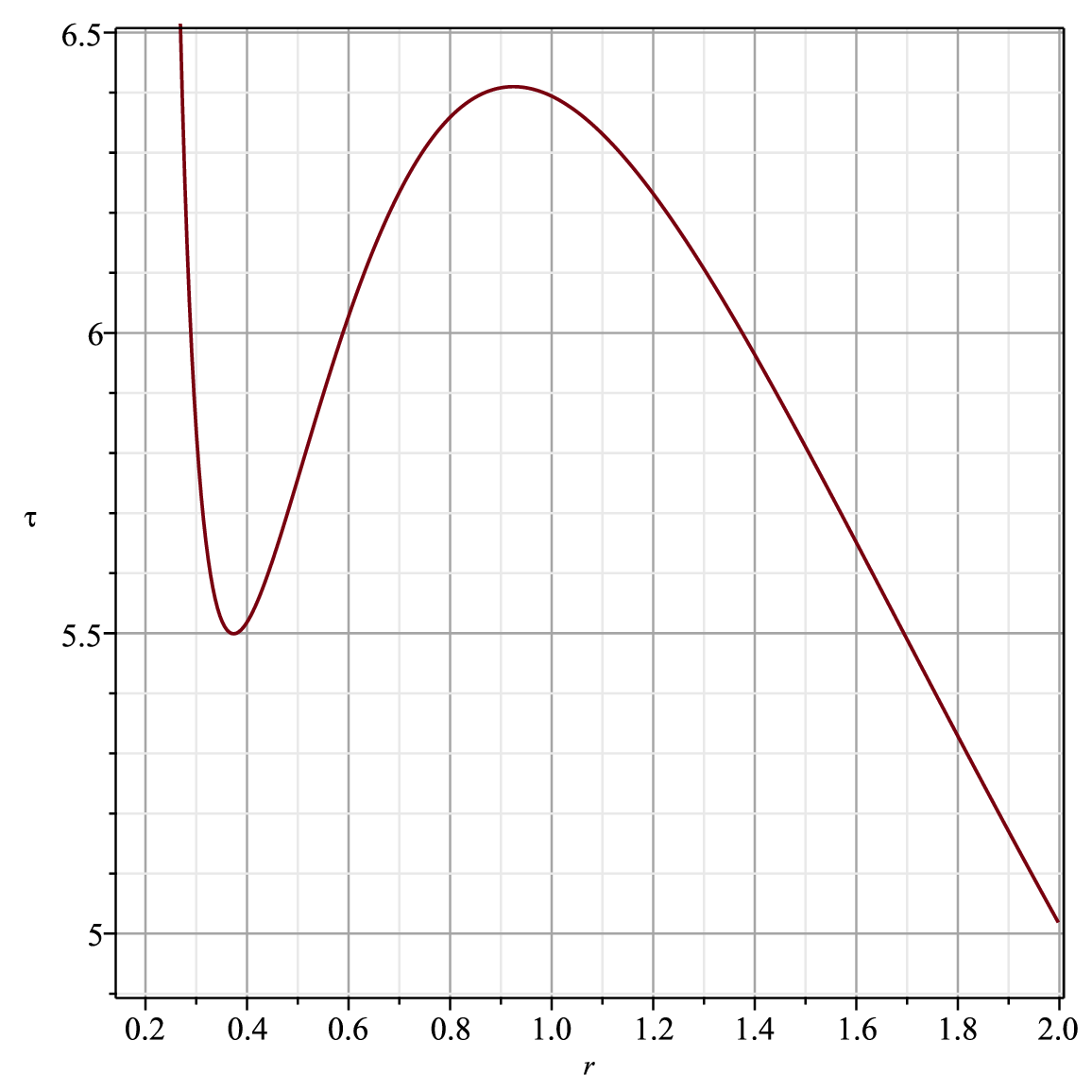}
 \label{1b}}
  \caption{\small{
The vector field $n$ on a part of the $(r-\Theta)$ plane for the AdS R-N black holes with $(q=1,G=0.04 < G_{critical}=0.0498)$ is shown by the blue arrows in Fig (1a). The ZP are at $(r,\theta)=(0.28,1.57), (0.58,1.57), (1.37,1.57)$. We choose three closed loops (purple loop) and (blue loop) that encircle the ZP. Fig (1b) shows the curve of equation (27).}}
 \label{1}
 \end{center}
 \end{figure}

\subsection{Case II}
In this subsection,the 4D AdS EGB black hole can be considered as\cite{114},
$$f(r)=1+\frac{r^2}{2\alpha}\bigg(1-\sqrt{1+4\alpha\bigg(\frac{2MG}{r^3}-\frac{q^2 G}{r^4}-\frac{1}{l^2}\bigg)}\bigg)$$
The radius of AdS and entropy for this case is given by,
\begin{equation}\label{28}
\begin{split}
&l = \frac{\sqrt{6}\, \sqrt{\frac{1}{P G \pi}}}{4},\\
&S  = \frac{r_h^{2} \pi}{G}+4 \ln \! \left(\frac{r_h}{\sqrt{\alpha}}\right) \alpha  \pi
\end{split}
\end{equation}
Also, the Hawking temperature of AdS EGB black holes is rewritten according to the equation (14) in the following form
\begin{equation}\label{29}
T =\frac{8 P G \pi  r_h^{4}-q^{2} G +r_h^{2}-\alpha}{4 r_h^{3} \pi +8 r_h \alpha  \pi}
\end{equation}
The variable cosmological constant for this model is expressed in the following form,
\begin{equation}\label{30}
G =-\frac{-r_h^{4}+5 \alpha  r_h^{2}+2 \alpha^{2}}{8 P \pi  r_h^{6}+48 P \pi  \alpha  r_h^{4}+3 q^{2} r_h^{2}+2 \alpha  q^{2}}
\end{equation}
Also, the mass and Helmholtz free energy for the mentioned black hole in BB framework are calculated as,
\begin{equation}\label{31}
M =\frac{r_h}{2 G}+\frac{\alpha}{2 G r_h}+\frac{4 r_h^{3} P \pi}{3}+\frac{q^{2}}{2 r_h}
\end{equation}
and
\begin{equation}\label{32}
\mathcal{F} =\frac{r_h}{2 G}+\frac{\alpha}{2 G r_h}+\frac{4 r_h^{3} P \pi}{3}+\frac{q^{2}}{2 r_h}-\frac{\frac{r_h^{2} \pi}{G}+4 \ln \! \left(\frac{r_h}{\sqrt{\alpha}}\right) \alpha  \pi}{\tau}
\end{equation}
The vector fields $\phi_{r}$ and $\phi_{\theta}$ are calculated as,
\begin{equation}\label{33}
\begin{split}
&\phi^{r_h} = \frac{\left(8 P G \pi  r_h^{4}-q^{2} G +r_h^{2}-\alpha \right) \tau -8 G \pi  \alpha  r_h -4 r_h^{3} \pi}{2 G \,r_h^{2} \tau},
\\
&\phi^{\Theta}=-\frac{\cos \! \left(\Theta \right)}{\sin \! \left(\Theta \right)^{2}}
\end{split}
\end{equation}
Like the previous case we can calculate the $\tau$,
\begin{equation}\label{34}
\tau =\frac{4 \pi  \left(2 G \alpha +r_h^{2}\right) r_h}{8 P G \pi  r_h^{4}-q^{2} G +r_h^{2}-\alpha}
\end{equation}

\begin{figure}[h!]
 \begin{center}
 \subfigure[]{
 \includegraphics[height=5.5cm,width=6cm]{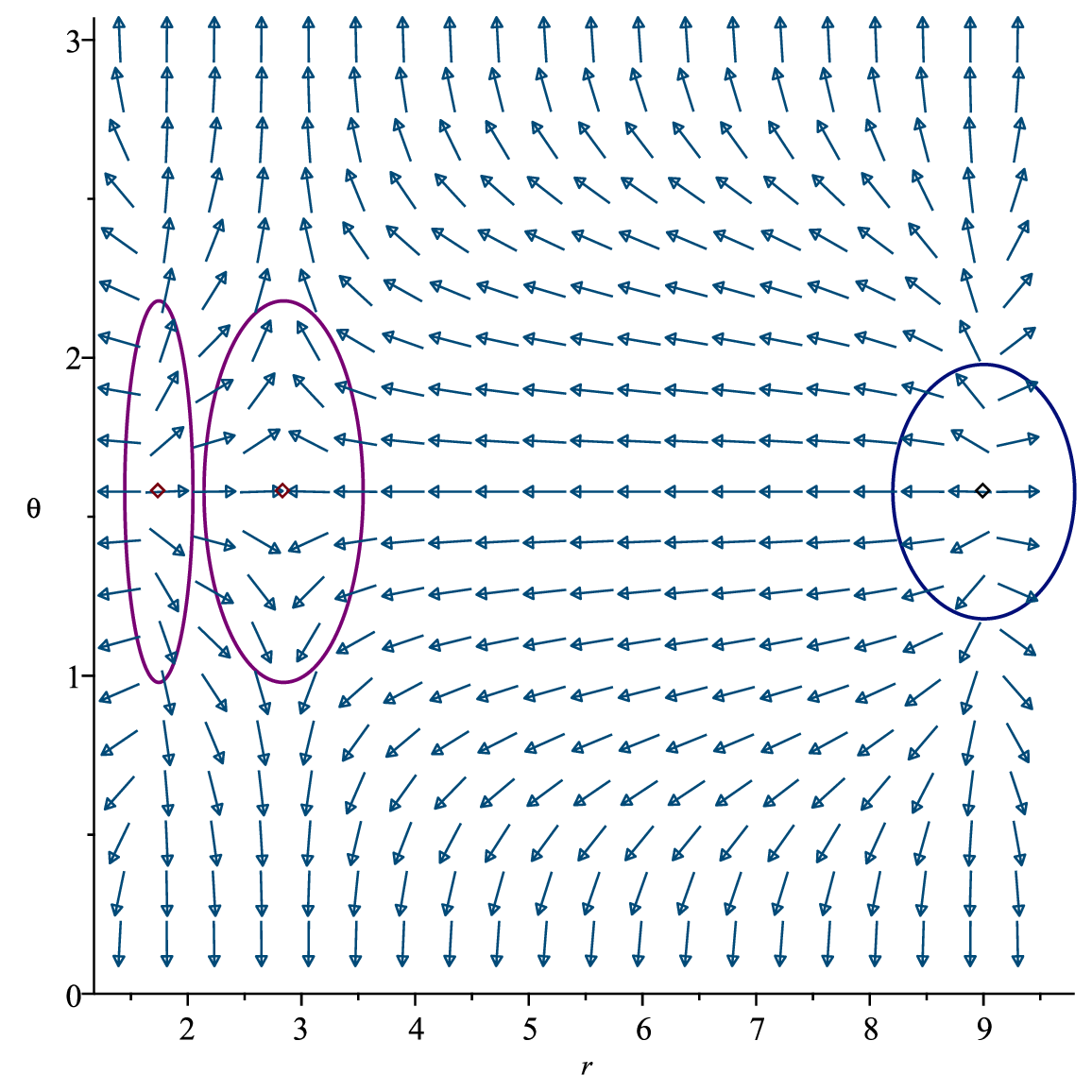}
 \label{2a}}
 \subfigure[]{
 \includegraphics[height=5.5cm,width=6cm]{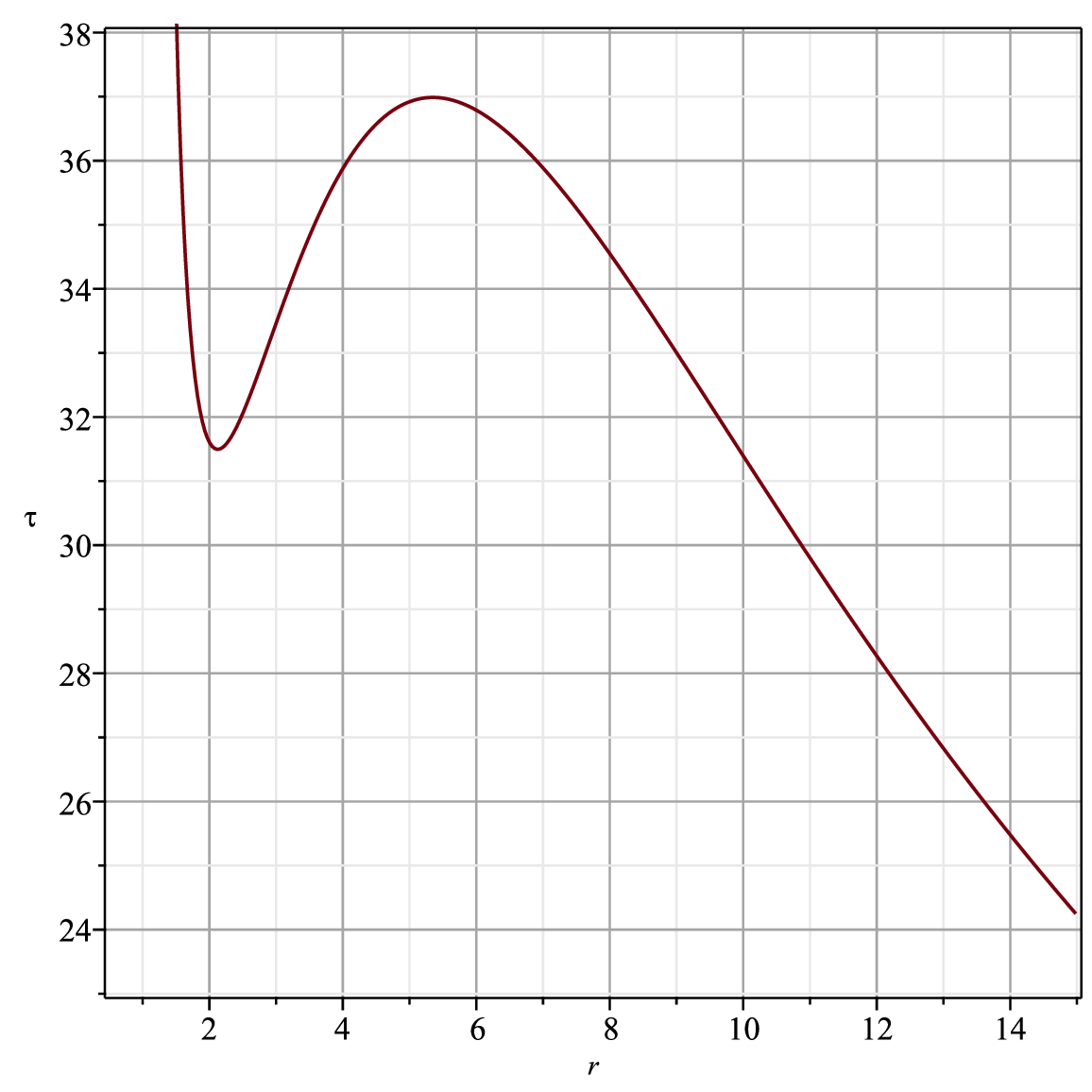}
 \label{2b}}
  \caption{\small{
The vector field $n$ on a part of the $(r-\Theta)$ plane for the AdS EGB black holes with $(P=0.1, G=0.012 < G_{critical}=0.01406567655,q=10, \alpha=0.1)$ is shown by the blue arrows in Fig (2a). The ZPs are at $(r,\theta)=(1.745506274,1.57), (2.842950665,1.57), (9.002657827,1.57)$. We choose three closed loops (purple loop) and (blue loop) that encircle the ZPs. Fig (2b) shows the curve of equation (34).}}
 \label{2}
 \end{center}
 \end{figure}

\subsection{Case III}
According to equation (18) for the AdS EPYM black holes can be considered as\cite{115}
$$f(r)=1-\frac{2GM}{r}+\frac{r^{2}}{l^2}+\frac{G(2q^2)^\gamma}{2(4\gamma-3)r^{4\gamma-2}}$$
Thermodynamic quantities such as Hawking temperature, mass, and Helmholtz free energy for AdS EPYM black holes are rewritten as follows
\begin{equation}\label{35}
T =\frac{1+8 r_h^{2} G P \pi -\frac{\left(2 q^{2}\right)^{\gamma} G}{2 r_h^{4 \gamma -2}}}{4 r_h \pi}
\end{equation}

\begin{equation}\label{36}
M =\frac{r_h \left(1+\frac{8 r_h^{2} G P \pi}{3}-\frac{q^{2 \gamma} 2^{-1+\gamma} G}{r_h^{4 \gamma -2} \left(4 \gamma -3\right)}\right)}{2 G}
\end{equation}
and Helmholtz free energy,
\begin{equation}\label{37}
\mathcal{F }=\frac{r_h \left(1+\frac{8 r_h^{2} G P \pi}{3}-\frac{q^{2 \gamma} 2^{-1+\gamma} G}{r_h^{4 \gamma -2} \left(4 \gamma -3\right)}\right)}{2 G}-\frac{r_h^{2} \pi}{G \tau}
\end{equation}
The $G$ which is given by,
\begin{equation}\label{38}
G =\frac{2}{8 \left(q^{2}\right)^{\gamma} r_h^{-4 \gamma +2} \gamma  2^{-1+\gamma}+16 r_h^{2} P \pi -\left(q^{2}\right)^{\gamma} r_h^{-4 \gamma +2} 2^{\gamma}}
\end{equation}
We calculate the two vector field $\phi_{r}$ and $\phi_{\theta}$,
\begin{equation}\label{39}
\phi^{r_h}=\frac{16 G \pi  P \,r_h^{2} \tau +G 2^{\gamma} q^{2 \gamma} r_h^{-4 \gamma +2} \tau -8 r_h \pi +2 \tau}{4 G \tau}
\end{equation}

\begin{equation}\label{40}
\phi^{\Theta}=-\frac{\cos \! \left(\Theta \right)}{\sin \! \left(\Theta \right)^{2}}
\end{equation}
Also the Euclidean time period $\tau$ is calculated as,
\begin{equation}\label{41}
\tau =\frac{8 r_h \pi}{16 r_h^{2} G P \pi +G 2^{\gamma} q^{2 \gamma} r_h^{-4 \gamma +2}+2}
\end{equation}

\begin{figure}[h!]
 \begin{center}
 \subfigure[]{
 \includegraphics[height=5.5cm,width=6cm]{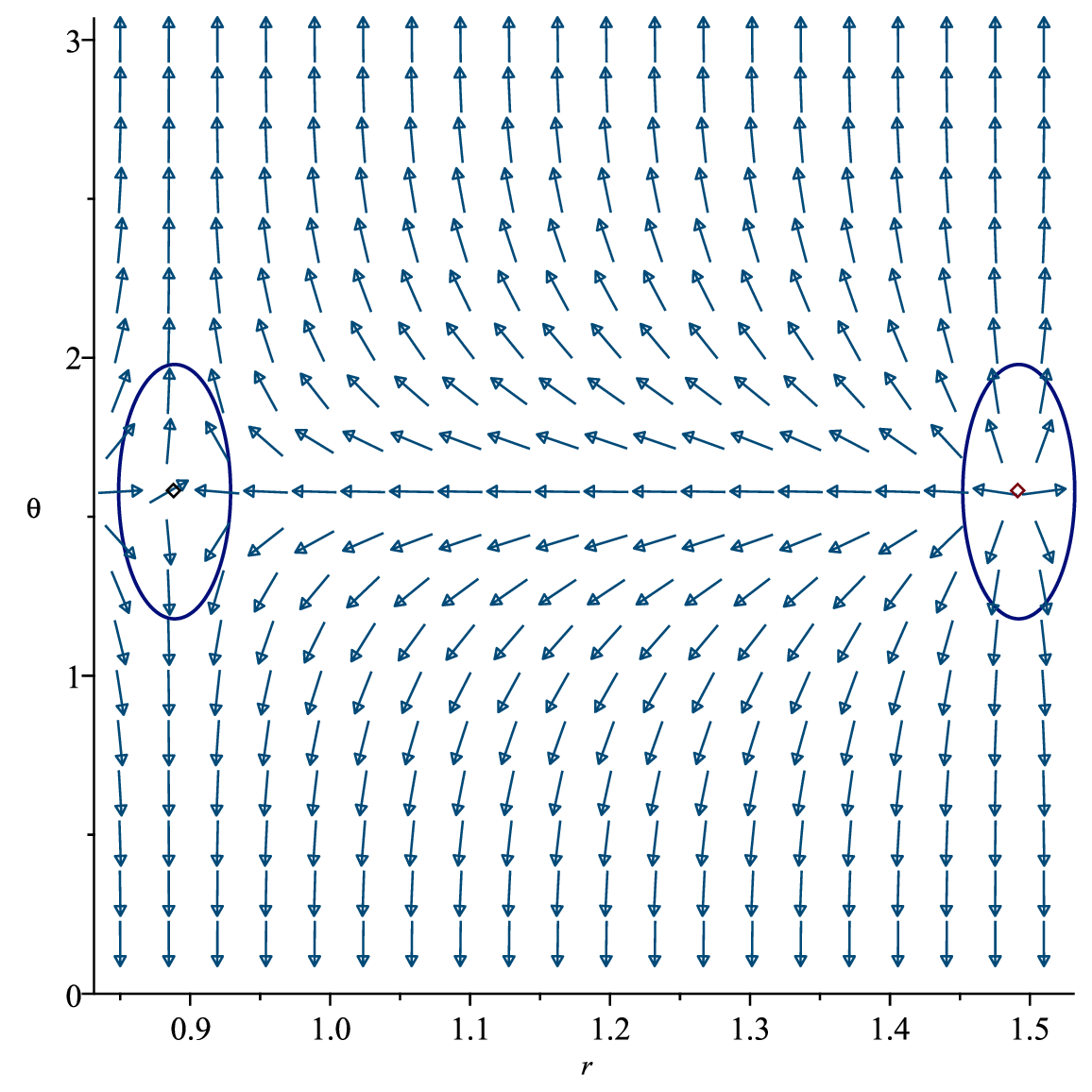}
 \label{3a}}
 \subfigure[]{
 \includegraphics[height=5.5cm,width=6cm]{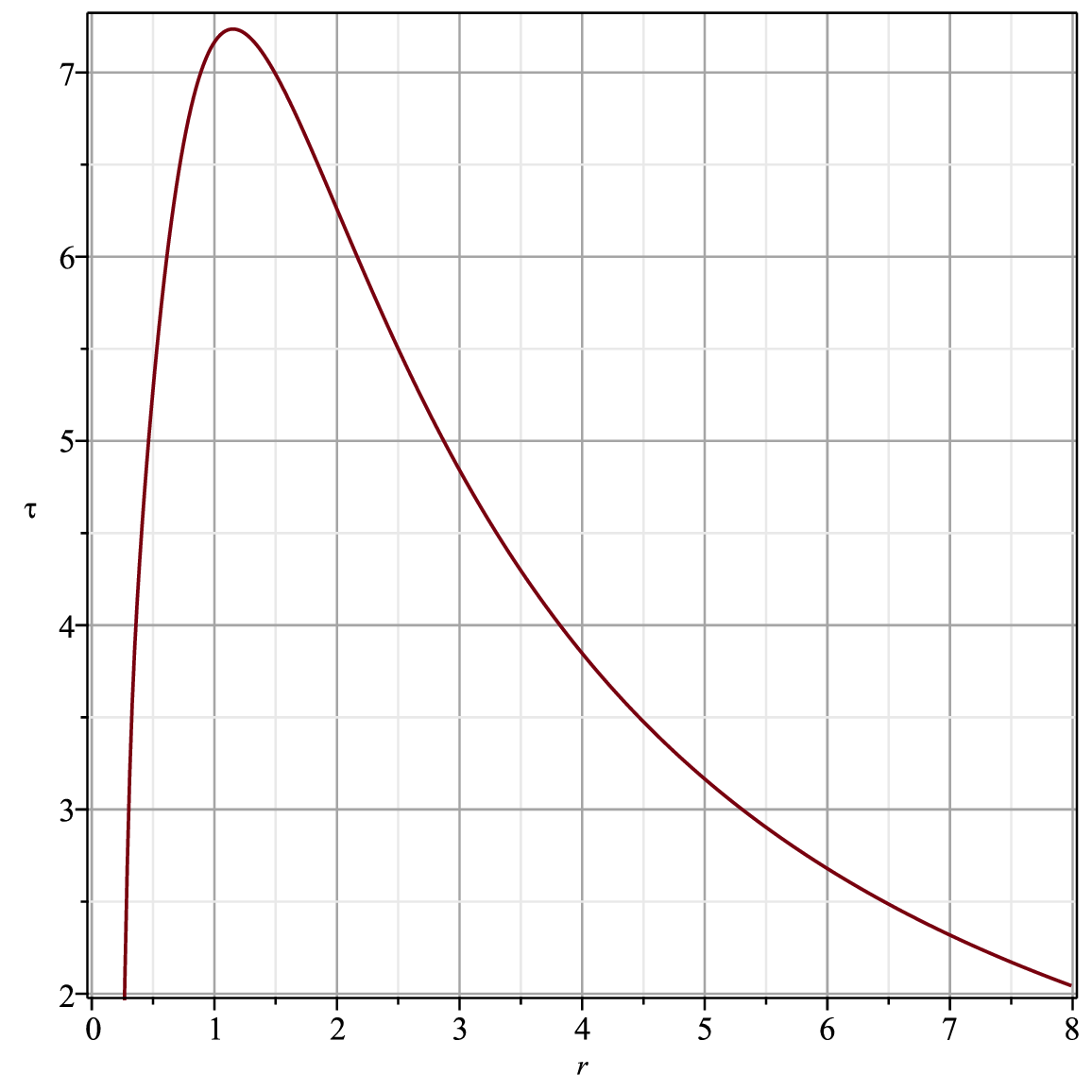}
 \label{3b}}
  \caption{\small{
The vector field $n$ on a part of the $(r-\Theta)$ plane for the AdS EPYM black holes with $( q=0.1, G=0.3 < G_{critical}=0.3068,P=0.1, \gamma=3)$ is shown by the blue arrows in Fig (3a). The ZPs are at $(r,\Theta)=(0.88,1.57), (1.49,1.57)$. We choose two closed loops (blue loop) that encircle the ZPs. Fig (3b) shows the curve of equation (41).}}
 \label{3}
 \end{center}
 \end{figure}

\section{Restricted Phase Space Thermodynamics}
RPS thermodynamics is a new formalism for thermodynamics of AdS black holes that is based on Visser's holographic thermodynamics but with the AdS radius fixed as a constant. RPST is free of the pressure and volume variables but inherits the central charge and chemical potential as a new pair of conjugate thermodynamic variables. RPST satisfies the Euler relation and the Gibbs-Duhem equation simultaneously with the first law of black hole thermodynamics, which guarantees the appropriate homogeneous behaviors for the black hole mass and the intensive variables. RPST reveals some interesting thermodynamic behaviors, such as supercritical phase transitions, novel critical points, and topological numbers. In this section, we aim to explore the thermodynamic topology of the three black holes mentioned above in the RPS. We explain the details of the calculations in the following subsections
\subsection{Case I}
In this subsection, we will rewrite the equations according to RPS, so entropy is rewritten with respect to equation (21) for AdS R-N black hole in the following form.
\begin{equation}\label{42}
\begin{split}
&q  = \frac{\hat{q}}{\sqrt{C}},\\
&G  = \frac{l^{2}}{C},\\
&S =\frac{C \,r_h^{2} \pi}{l^{2}}
\end{split}
\end{equation}
Therefore, the temperature of this black hole is given by,
\begin{equation}\label{43}
T =\frac{C^{2} l^{2} r_h^{2}+3 C^{2} r_h^{4}-l^{4} \hat{q}^{2}}{4 \pi  l^{3} \sqrt{\frac{C^{2} r_h^{2}}{l^{2}}}\, C \,r_h^{2}}
\end{equation}
We can obtain the parameter $C$ according to the mentioned equations, which is calculated as,
\begin{equation}\label{44}
C =\frac{3 l^{2} \hat{q}}{\sqrt{3 l^{2}-9 r_h^{2}}\, r_h}
\end{equation}
Also, we can obtain the critical value ($r_{c}=\frac{\sqrt{6}l}{6}$), so we can compute thermodynamic quantities such as mass and Helmholtz free energy,
\begin{equation}\label{45}
M =\frac{C^{2} l^{2} r_h^{2}+C^{2} r_h^{4}+l^{4} \hat{q}^{2}}{2 l^{5} \sqrt{\frac{C^{2} r_h^{2}}{l^{2}}}}
\end{equation}
and
\begin{equation}\label{46}
\mathcal{F }=\frac{C^{2} l^{2} r_h^{2}+C^{2} r_h^{4}+l^{4} \hat{q}^{2}}{2 l^{5} \sqrt{\frac{C^{2} r_h^{2}}{l^{2}}}}-\frac{C \,r_h^{2} \pi}{l \tau}
\end{equation}
Therefore, two field vectors ($\phi^{r_h}$) and ($\phi_{\Theta}$) are obtained
\begin{equation}\label{47}
\begin{split}
&\phi^{r_h} = \frac{-4 C \pi  l^{4} r_h^{2} \sqrt{\frac{C^{2} r_h^{2}}{l^{2}}}+\tau  \left(C^{2} l^{2} r_h^{2}+3 C^{2} r_h^{4}-l^{4} \hat{q}^{2}\right)}{2 \sqrt{\frac{C^{2} r_h^{2}}{l^{2}}}\, r_h \,l^{5} \tau},
\\
&\phi^{\Theta}=-\frac{\cos \! \left(\Theta \right)}{\sin \! \left(\Theta \right)^{2}}
\end{split}
\end{equation}
Hence we will have
\begin{equation}\label{48}
\tau =\frac{4 C \pi  l^{4} r_h^{2} \sqrt{\frac{C^{2} r_h^{2}}{l^{2}}}}{C^{2} l^{2} r_h^{2}+3 C^{2} r_h^{4}-l^{4} \hat{q}^{2}}
\end{equation}

\begin{figure}[h!]
 \begin{center}
 \subfigure[]{
 \includegraphics[height=5.5cm,width=6cm]{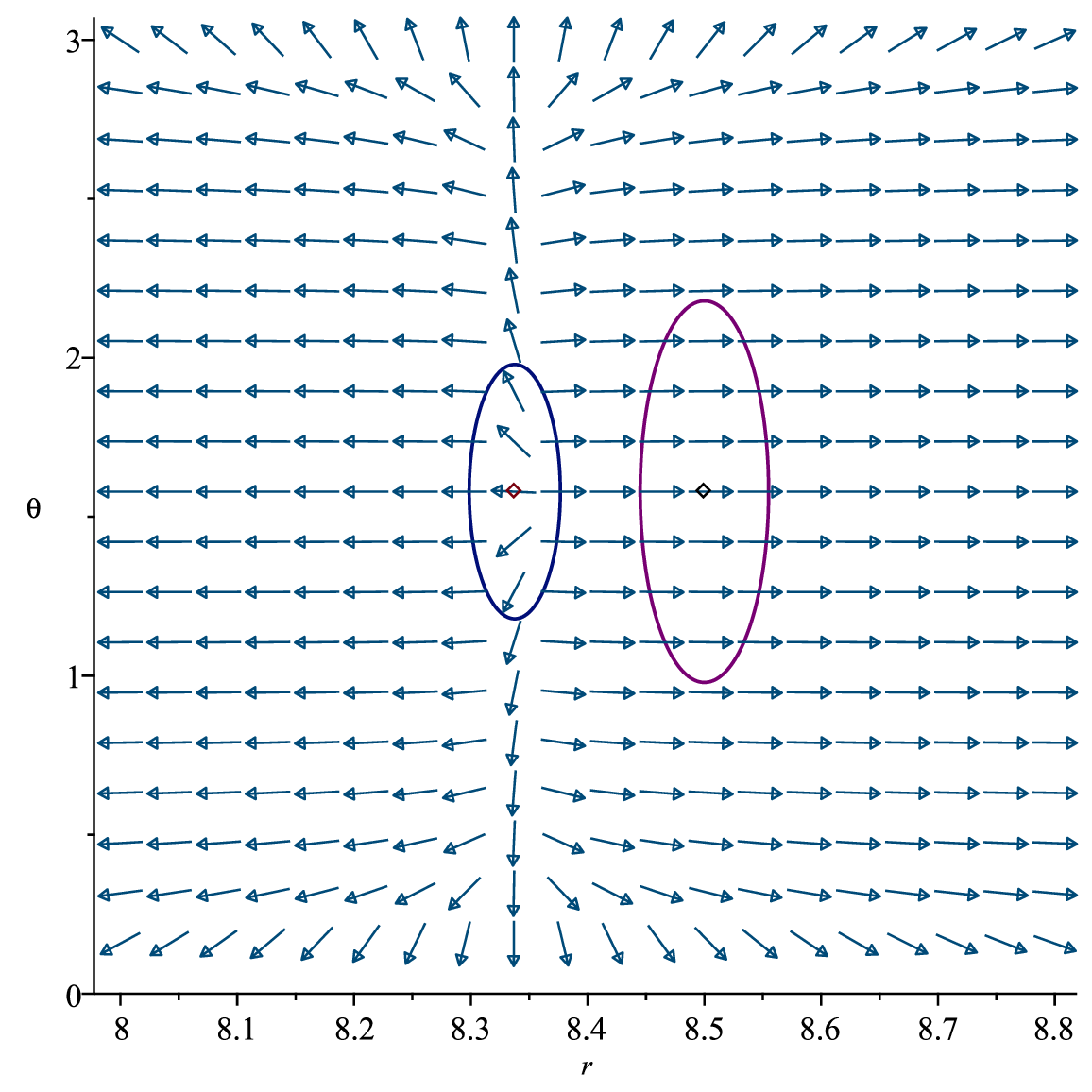}
 \label{4a}}
 \subfigure[]{
 \includegraphics[height=5.5cm,width=6cm]{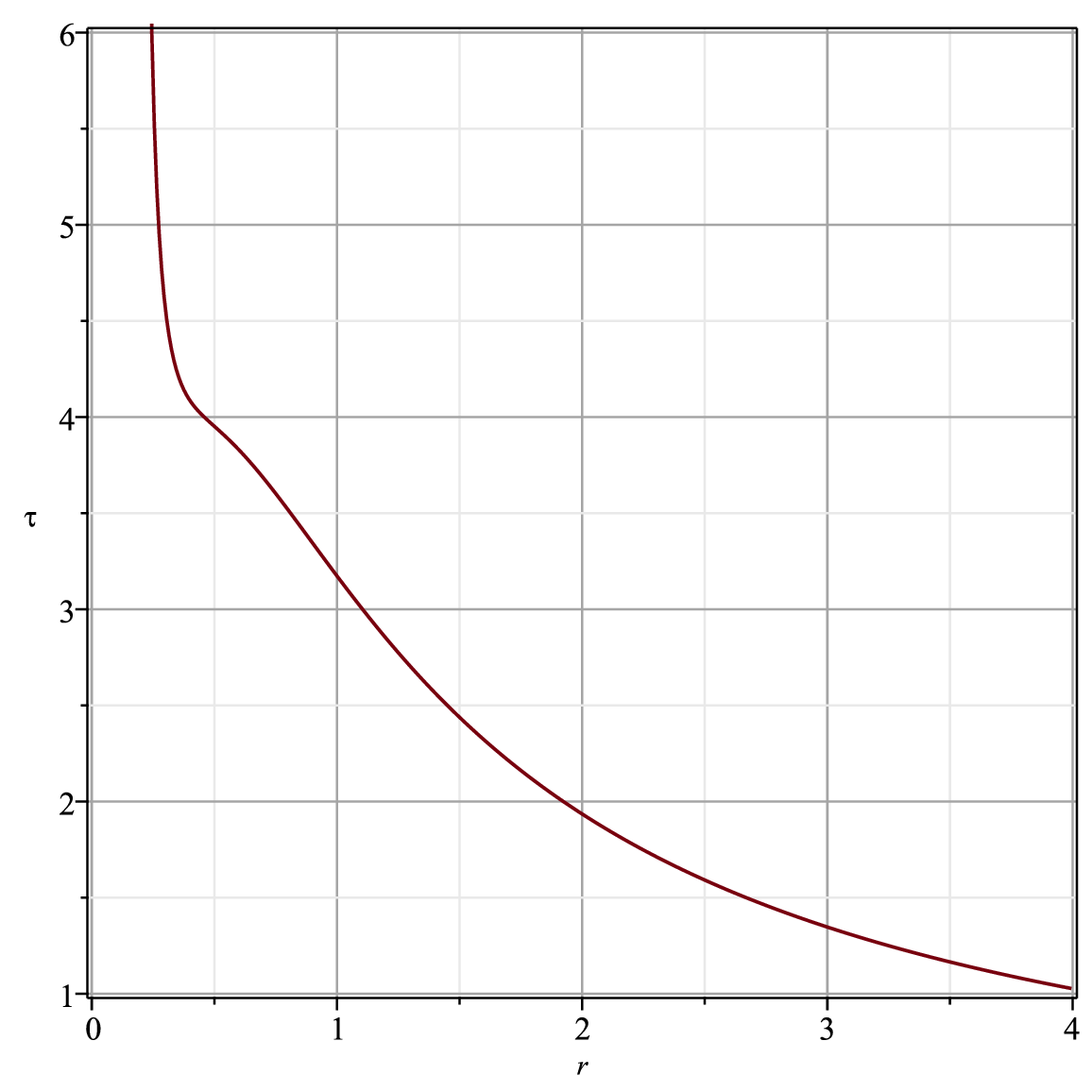}
 \label{4b}}
  \caption{\small{
The vector field $n$ on a part of the $(r-\Theta)$ plane for the AdS R-N black holes with $(\hat{q}=1, l=1,C=5 < C_{critical}=6)$ is shown by the blue arrows in Fig (4a). The ZP is at $(r,\Theta)=(8.37,1.57)$. We choose two closed loops $Z_1$ (purple loop) and $Z_2$ (blue loop), where $Z_2$ encircles the ZP but $Z_1$ does not. Fig (4b) shows the curve of equation (48).}}
 \label{4}
 \end{center}
 \end{figure}

\subsection{Case II}
Here, we will rewrite the equations like pervious parts for A4D AdS EGB black hole. So entropy is rewritten with respect to equation (28) in the following form.
\begin{equation}\label{49}
\begin{split}
&q  = \frac{\hat{q}}{\sqrt{C}},\\
&S  = \frac{C \,r_h^{2} \pi}{l^{2}}+4 \ln \! \left(\frac{r_h}{\sqrt{\alpha}}\right) \alpha  \pi ,\\
&G  = \frac{l^{2}}{C}
\end{split}
\end{equation}

Thus, the temperature (T) is given by,
\begin{equation}\label{50}
T =\frac{-\frac{\hat{q}^{2} l^{4}}{C^{2}}+l^{2} r_h^{2}+3 r^{4}-l^{2} \alpha}{4 \left(r_h^{2}+2 \alpha \right) r_h \,l^{2} \pi}
\end{equation}
The parameter $C$ which is calculated as,
\begin{equation}\label{51}
C =\frac{\sqrt{-\left(-l^{2} r_h^{4}+3 r_h^{6}+5 l^{2} \alpha  r_h^{2}+18 \alpha  r_h^{4}+2 \alpha^{2} l^{2}\right) \left(3 r_h^{2}+2 \alpha \right)}\, l^{2} \hat{q}}{-l^{2} r_h^{4}+3 r_h^{6}+5 l^{2} \alpha  r_h^{2}+18 \alpha  r_h^{4}+2 \alpha^{2} l^{2}}
\end{equation}
Also, we can compute thermodynamic quantities such as mass and Helmholtz free energy as follows,
\begin{equation}\label{52}
M =\frac{\left(\left(r_h^{2}+\alpha \right) l^{2}+r_h^{4}\right) C^{2}+\hat{q}^{2} l^{4}}{2 l^{4} r_h C}
\end{equation}
\begin{equation}\label{53}
\mathcal{F }=\frac{\left(\left(r_h^{2}+\alpha \right) l^{2}+r_h^{4}\right) C^{2}+\hat{q}^{2} l^{4}}{2 l^{4} r_h C}-\frac{\frac{C \,r_h^{2} \pi}{l^{2}}+4 \ln \! \left(\frac{r_h}{\sqrt{\alpha}}\right) \alpha  \pi}{\tau}
\end{equation}
The vector field vectors $\phi_{r}$ is obtained as,
\begin{equation}\label{54}
\phi^{r_h}=\frac{\left(\left(\left(r_h^{2}-\alpha \right) \tau -4 r_h^{3} \pi \right) l^{2}+3 r_h^{4} \tau \right) C^{2}-8 C \pi  \alpha  l^{4} r_h -l^{4} \tau  \hat{q}^{2}}{2 l^{4} r_h^{2} C \tau}
\end{equation}
Hence we will have,
\begin{equation}\label{55}
\tau =-\frac{4 \pi  \left(C \,r_h^{2}+2 l^{2} \alpha \right) l^{2} r_h C}{-r^{2} C^{2} l^{2}-3 r_h^{4} C^{2}+\hat{q}^{2} l^{4}+\alpha  C^{2} l^{2}}
\end{equation}

\begin{figure}[h!]
 \begin{center}
 \subfigure[]{
 \includegraphics[height=5.5cm,width=6cm]{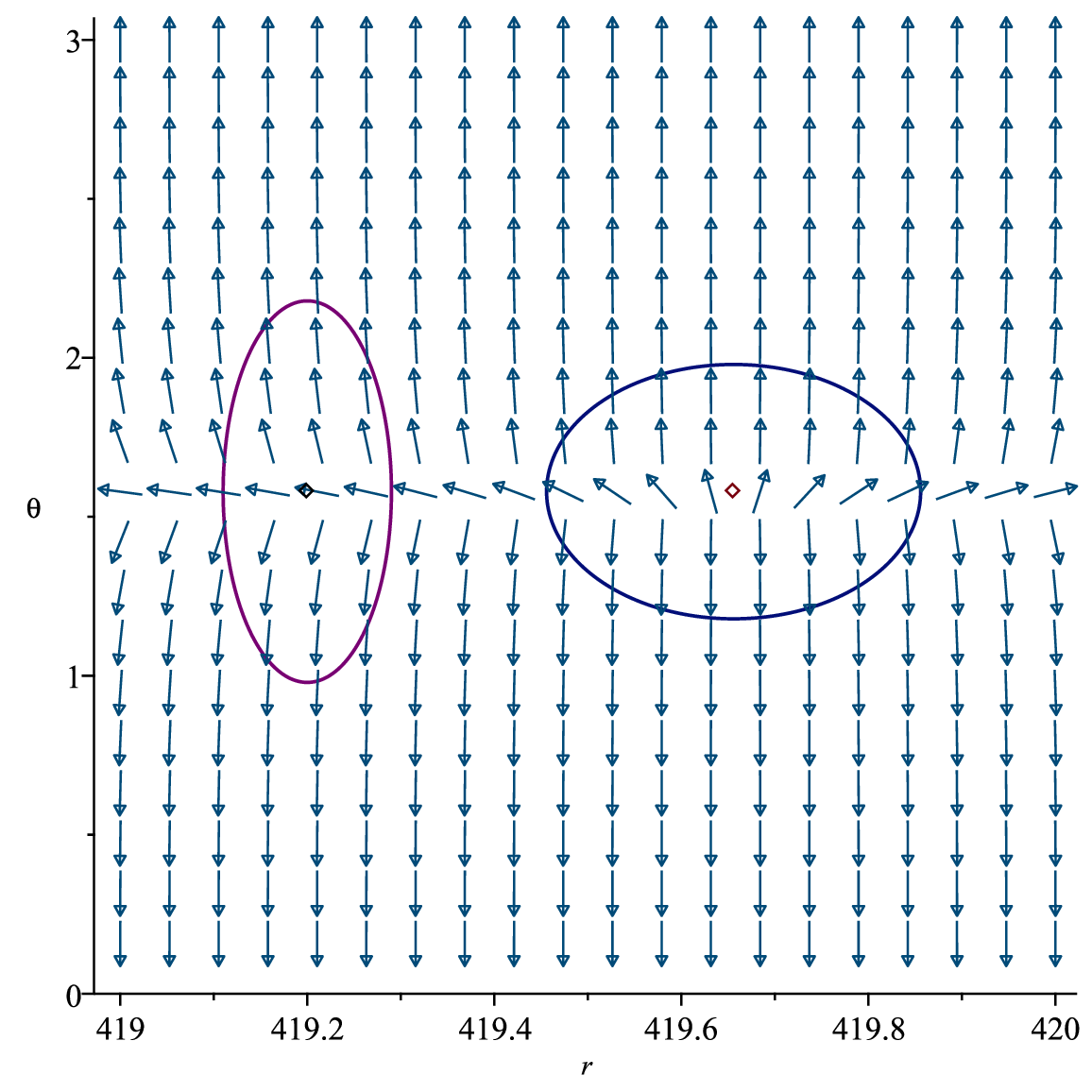}
 \label{5a}}
 \subfigure[]{
 \includegraphics[height=5.5cm,width=6cm]{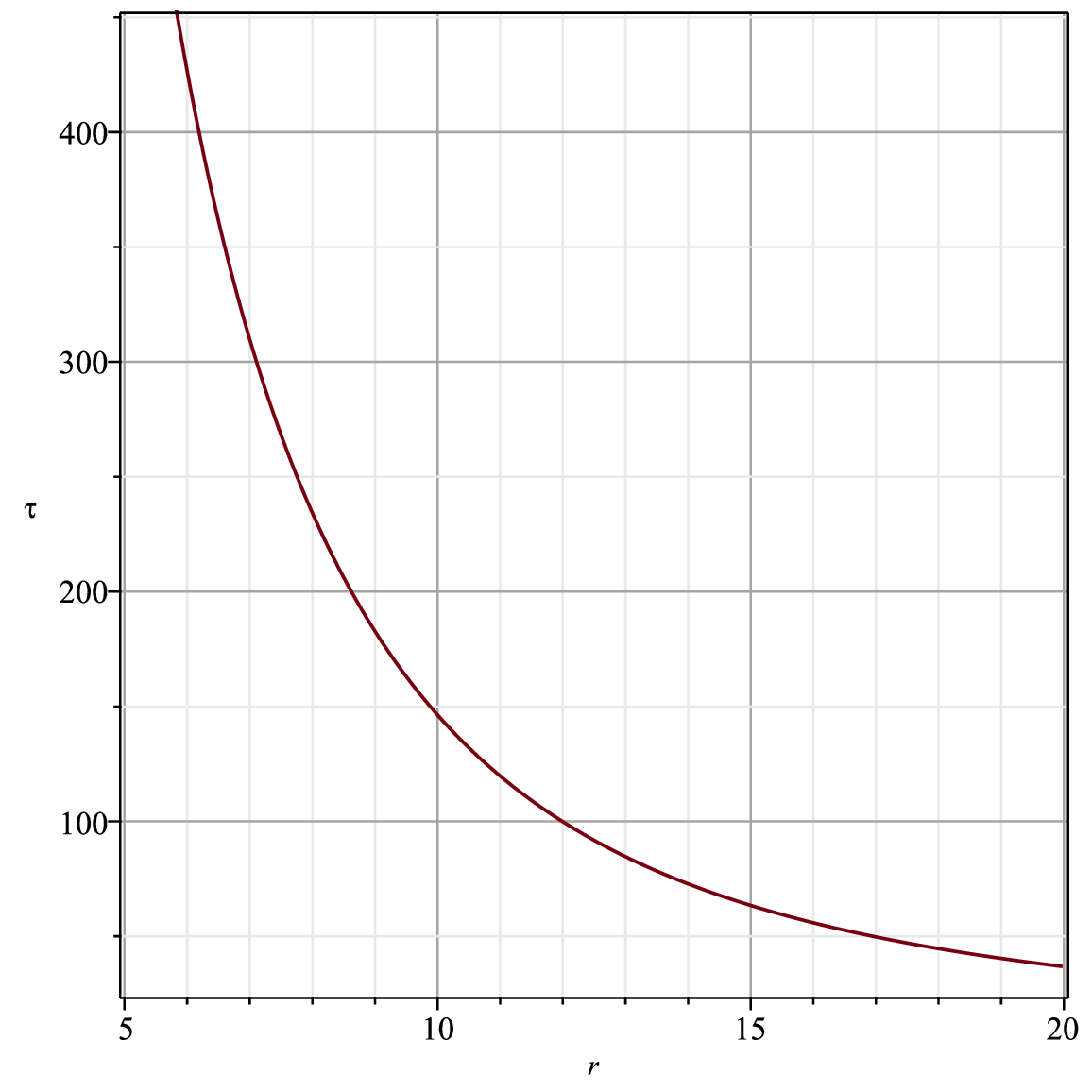}
 \label{5b}}
  \caption{\small{
The vector field $n$ on a part of the $(r-\Theta)$ plane for the AdS EGB black holes with $(C=0.5 < C_{critical}=0.9447273930,\hat{q}=0.1, \alpha=0.9, l=10)$ is shown by the blue arrows in Fig (5a). The ZPs are at $(r,\Theta)=(419.6558504,1.57)$. We choose two closed loops $Z_1$ (purple loop) and $Z_2$ (blue loop), where $Z_2$ encircles the ZPs but $Z_1$ does not. Fig (5b) shows the curve of equation (55).}}
 \label{5}
 \end{center}
 \end{figure}

\subsection{Case III}
Here, we will rewrite the equations for AdS EPYM black holes. The entropy for this black hole is as follows,
\begin{equation}\label{56}
\begin{split}
&q  = \frac{\hat{q}}{\sqrt{C}},\\
&G  = \frac{l^{2}}{C},\\
&S  = \frac{C \,r_h^{2} \pi}{l^{2}}
\end{split}
\end{equation}
Therefore, the temperature of AdS EPYM black holes in RPS thermodynamics is given by,
\begin{equation}\label{57}
T =\frac{1+\frac{r_h^{2}}{l^{2}}-\frac{\left(\frac{2 \hat{q}^{2}}{C}\right)^{\gamma} l^{2}}{2 C \,r_h^{4 \gamma -2}}}{4 r_h \pi}
\end{equation}
The parameter $C$ according to the mentioned equations, which is obtained as,
\begin{equation}\label{58}
C=2 \hat{q}^{2} \exp\bigg({-\frac{2 \ln \left(2\right)-\ln \left(4 \gamma -1\right)+\ln \left(\frac{\hat{q}^{2} r_h^{4 \gamma -2} \left(l^{2}-r_h^{2}\right)}{l^{4}}\right)}{\gamma +1}}\bigg)
\end{equation}
We compute the mass and Helmholtz free energy for this black hole,
\begin{equation}\label{59}
M =\frac{r_h \left(1+\frac{r_h^{2}}{l^{2}}-\frac{\left(\frac{\hat{q}}{\sqrt{C}}\right)^{2 \gamma} 2^{\gamma -1} l^{2}}{C \,r_h^{4 \gamma -2} \left(4 \gamma -3\right)}\right) C}{2 l^{2}}
\end{equation}
and
\begin{equation}\label{60}
\mathcal{F} =\frac{r_h \left(1+\frac{r_h^{2}}{l^{2}}-\frac{\left(\frac{\hat{q}}{\sqrt{C}}\right)^{2 \gamma} 2^{\gamma -1} l^{2}}{C \,r_h^{4 \gamma -2} \left(4 \gamma -3\right)}\right) C}{2 l^{2}}-\frac{C \,r_h^{2} \pi}{l^{2} \tau}
\end{equation}
Therefore, $(\phi_{r})$ is obtained as,
\begin{equation}\label{61}
\phi^{r_h}=\frac{\left(1+\frac{r_h^{2}}{l^{2}}-\frac{\left(\frac{\hat{q}}{\sqrt{C}}\right)^{2 \gamma} 2^{\gamma -1} l^{2}}{C \,r_h^{4 \gamma -2} \left(4 \gamma -3\right)}\right) C}{2 l^{2}}+\frac{r_h \left(\frac{2 r_h}{l^{2}}+\frac{\left(\frac{\hat{q}}{\sqrt{C}}\right)^{2 \gamma} 2^{\gamma -1} l^{2} \left(4 \gamma -2\right)}{C \,r^{4 \gamma -2} \left(4 \gamma -3\right) r_h}\right) C}{2 l^{2}}-\frac{2 C r_h \pi}{l^{2} \tau}
\end{equation}
Also, we can calculate the $\phi_{\theta}$,
\begin{equation}\label{62}
\phi^{\Theta}=-\frac{\cos \! \left(\Theta \right)}{\sin \! \left(\Theta \right)^{2}}
\end{equation}
So, We will have,
\begin{equation}\label{63}
\tau =\frac{4 C r_h \pi  l^{2} r^{4 \gamma -2}}{\left(\frac{\hat{q}}{\sqrt{C}}\right)^{2 \gamma} 2^{\gamma -1} l^{4}+C \,r_h^{4 \gamma -2} l^{2}+3 C \,r_h^{4 \gamma -2} r_h^{2}}
\end{equation}

\begin{figure}[h!]
 \begin{center}
 \subfigure[]{
 \includegraphics[height=5cm,width=5cm]{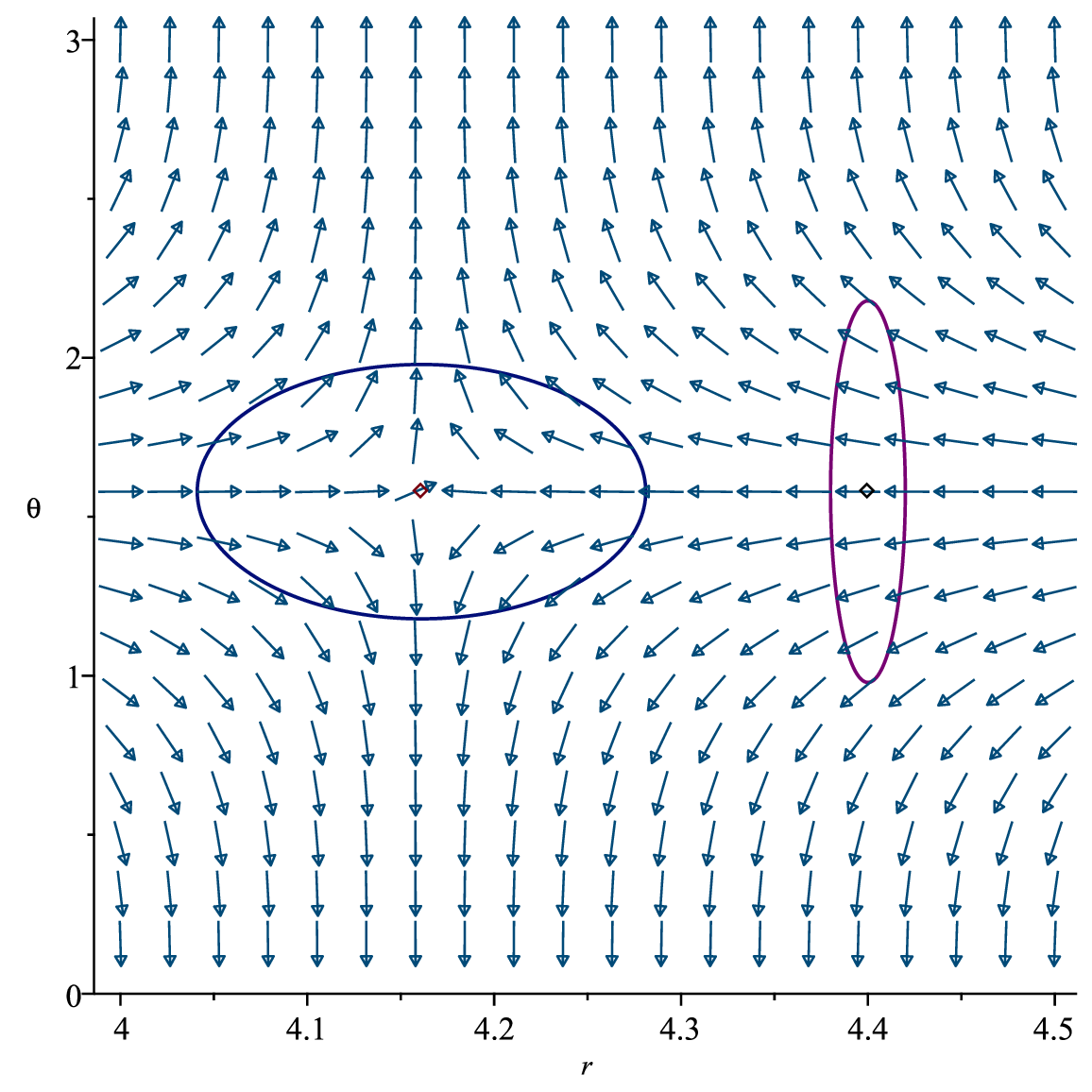}
 \label{6a}}
 \subfigure[]{
 \includegraphics[height=5cm,width=5cm]{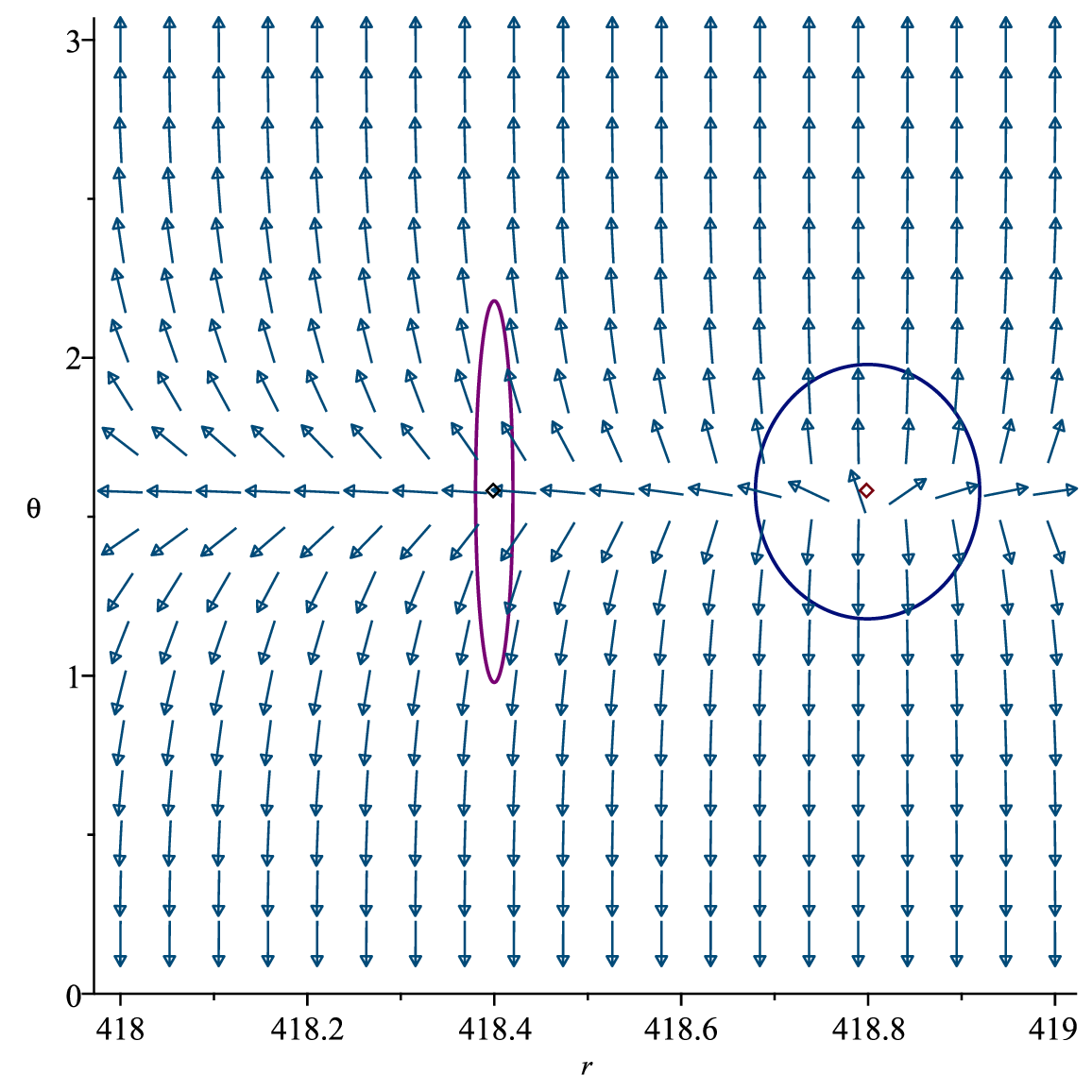}
 \label{6b}}
 \subfigure[]{
 \includegraphics[height=5cm,width=5cm]{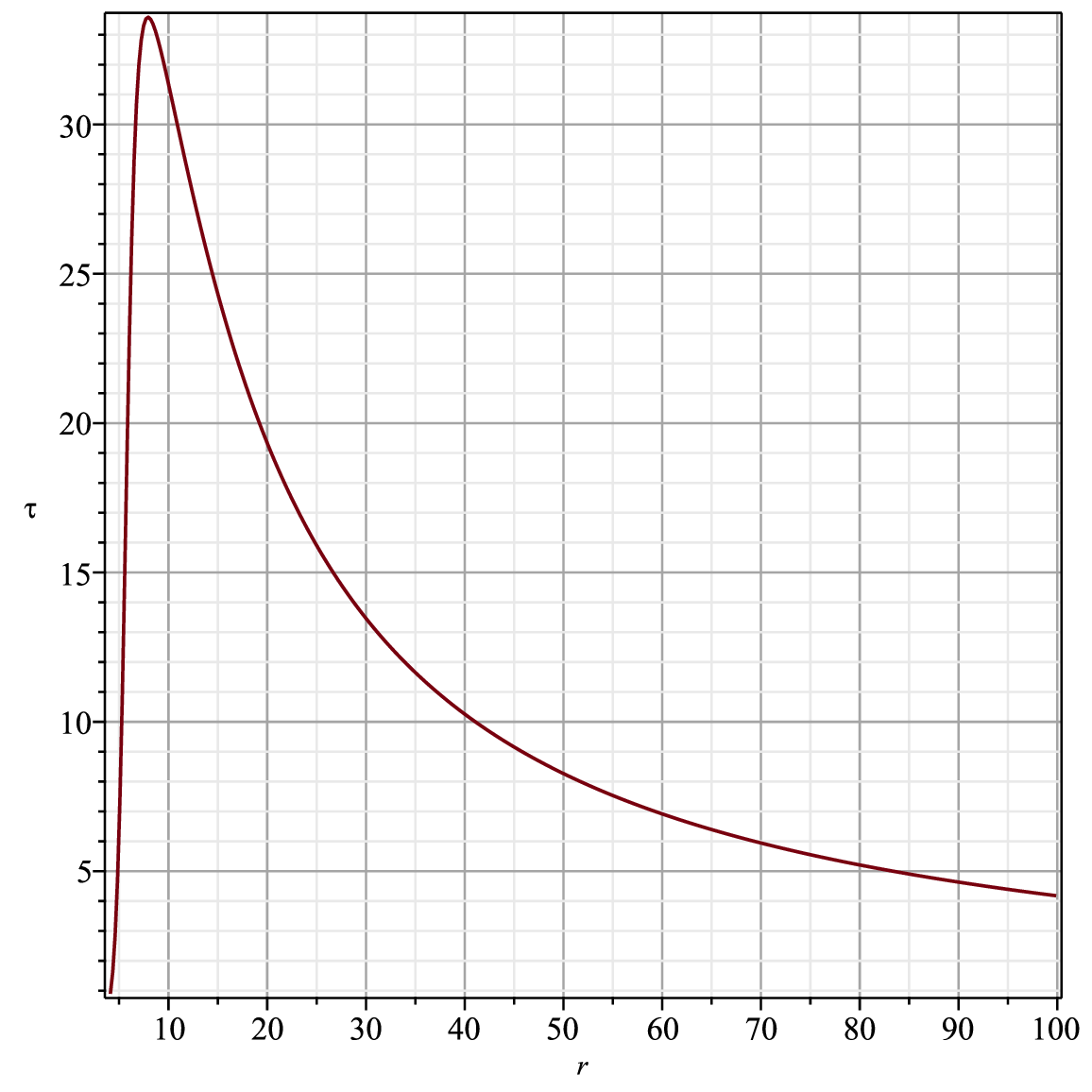}
 \label{6c}}
  \caption{\small{
The vector field $n$ on a part of the $(r-\Theta)$ plane for the AdS EPYM black holes with $(\hat{q}=10,C=1.5 < C_{critical}=1.6, l=10, \gamma=3,)$ is shown by the blue arrows in Fig (6a). The ZPs are at $(r,\Theta)=(4.16,1.57), (418,1.57)$. We choose four closed loops $Z_1$ (purple loop) and (blue loop), where (blue loop) encircles the ZPs but $Z_1$ does not. Fig (6c) shows the curve of equation (63).}}
 \label{6}
 \end{center}
 \end{figure}
 We will perform more calculations and discuss the results and analyze the figures in section Appendix A to check whether the dimension has an effect on the thermodynamic topology in both spaces or not, and also to check whether the thermodynamic topological behavior of 4D EPYM black hole in other spaces such as EPST is similar to the BBT and RPS
\section{Discussion and Result}
Here, we analyze Figures (1-6), which relate to three black holes, namely AdS R-N, AdS EGB, and AdS EPYM.
Considering that the general structure of the description of the shapes follows the same structural pattern, also, with respect to the numbers of the shapes, to avoid confusion and better understanding, we will express the pattern of interpretation and classify it for a model. The rest will follow the same pattern. In general, when the contour includes a winding:\\

1. If the field lines diverge around the zero point (like positive electric charges), its topological charge is +1.\\

2. If the field lines converge around the zero point (like negative electric charges), its topological charge is -1.\\

3. If the contour does not include a winding, its charge will be zero.\\

Finally, the total charges for each shape are the algebraic sum of the charges, which can be displayed by drawing a contour that includes all the zero points.\\
Figures (1) and (4) include zero points (ZPs) related to the AdS R-N black hole in the bulk boundary and RPS, respectively. They are divided into two parts. The plots (1a) and (4a) exhibit the normalized field lines. As illustrated in Figure (1a), there are three topological numbers $(\omega=+1, \omega=-1, \omega=+1)$ with $W=1$, but Figure 4a has only one ZP, which represents one topological number (+1) for the AdS R-N black hole in the RPS, i.e., the total topological number is $W=1$. They are proportional to the winding number and ZPs are located inside the contours at the coordinates $(r, \theta) = (0.28, 1.57); (0.58, 1.57); (1.37, 1.57)$ for the bulk boundary in Fig (1a) and $(r, \theta) = (8.37, 1.57)$ for the RPS in Fig 4a with respect to the free parameters that are mentioned in the figures. In Figs (1b) and (4b), we plot the curves related to equations (27) and (48) for the mentioned free parameters. As shown in Fig (1b), there are three black holes with respect to the maximum and minimum points which are related to one small black hole branch and one large black hole branch for the bulk boundary frame, but in Fig (4b) there is only one on-shell BH for arbitrary values of $(\tau)$ for the AdS R-N black hole in the RPS. The results for the AdS EGB black holes are exactly the same as those for the AdS R-N black hole. We also plotted normalized field lines for the AdS EPYM black holes in the bulk boundary in Fig (3a) that have topological numbers of $(\omega=+1, \omega=-1)$ with $W=0$. Also, we have an exactly similar behavior for this black hole in the RPS i.e., in Figs (6a) and (6b) we have the topological number $(\omega=+1, \omega=-1)$ with $W=0$. Because ZPs are far apart, we had to use two different plots to display them, i.e. (6a) and (6b) so that ZPs can be seen better. This similar behavior in both spaces is very attractive to us, which we will explain in detail later. Unlike the previous two black holes, AdS EPYM black hole has the same plots for $r-\tau$ in both spaces i.e., Figs (3b) and (6b) have the same behaviors. As $\tau$ decreases monotonically with the horizon radius $r$, it can be concluded that there is only one on-shell black hole for an arbitrarily fixed $\tau$, and no phase transition occurs. As shown in Figs (3b, 4b, 5b and 6b), there is only one on-shell black hole for arbitrary values of ($\tau$).\\
Before finishing the discussion, it is necessary to explain that the annihilation and generation points are very interesting and important for the time evolution of a black hole. They indicate the possible changes in the phase structure and stability of the black hole as it interacts with its environment.
To find these points, it is enough to check the following conditions:\\

For generation points we should have
\begin{equation}\label{(1)}
\frac{d}{d r_{h}}\tau \! \left(r_{h}\right)=0,\hspace{0.3cm} 0<\frac{d^{2}}{d r_{h}^{2}}\tau \! \left(r_{h}\right)
\end{equation}
and for annihilation points we should have
\begin{equation}\label{(2)}
\frac{d}{d r_{h}}\tau \! \left(r_{h}\right)=0,\hspace{0.3cm}  \frac{d^{2}}{d r_{h}^{2}}\tau \! \left(r_{h}\right)<0
\end{equation}

We summarize our results in Tables 1, 2 and 3.

\begin{center}
\begin{table}
  \centering
 \begin{tabular}{|p{4.5cm}|p{3.5cm}|p{5cm}|}
   \hline
   \centering{Case} & \centering{Thermodynamics} & Topological Number  \\
 \hline
  \multirow{3}{4.5cm}{}&\centering{$RPST$} & $\omega=+1$, $W=+1$  \\[2mm]
   AdS R-N Black Hole&\centering{$BBT$} & $\omega=+1, -1, +1$, $W=+1$ \\[2mm]
   \hline
    \multirow{3}{4.5cm}{}& \centering{$RPST$} & $\omega=+1$, $W=+1$   \\[2mm]
     AdS EGB Black Hole& \centering{$BBT$}& $\omega=+1, -1 +1$, $W=+1$  \\[2mm]
   \hline
   \multirow{3}{4.5cm}{}&\centering{$RPST$} & $\omega=-1, +1$, $W=0$  \\[2mm]
     AdS EPYM Black Hole& \centering{$BBT$}& $\omega=-1, +1$, $W=0$  \\[2mm]
   \hline
 \end{tabular}
\caption{Summary of the results. }\label{1}
\end{table}
 \end{center}

 \begin{center}
\begin{table}
\centering
 \begin{tabular}{|c|p{5cm}|p{5cm}|}
   \hline
   Case &Generation Point &Annihilation Point\\
   \hline
   AdS R-N Black Hole & BBT =1 & BBT =1\\
   \hline
   AdS EGB Black Hole & BBT =1 & BBT =1 \\
   \hline
   AdS EPYM Black Hole & BBT =0 & BBT =1\\
   \hline
 \end{tabular}
\caption{The generation and annihilation points of BBT}\label{2}
\end{table}
 \end{center}
\vspace{-2cm}
 \begin{center}
\begin{table}
\centering
 \begin{tabular}{|c|p{5cm}|p{5cm}|}
   \hline
   Case &Generation Point &Annihilation Point\\
   \hline
   AdS R-N Black Hole & RPS =0 &  RPS =0\\
   \hline
   AdS EGB Black Hole & RPS =0 &  RPS =0\\
   \hline
   AdS EPYM Black Hole & RPS =0 &  RPS =1\\
   \hline
 \end{tabular}
\caption{The generation and annihilation points of RPS}\label{2}
\end{table}
 \end{center}
\newpage
\section{Concluding Remark}
In this article, we investigated the thermodynamic topology of three black holes, namely AdS R-N, AdS EGB, and AdS EPYM, from two different frameworks: BB and RPS. Using the generalized off-shell Helmholtz free energy method, we calculated the thermodynamic topology of the selected black holes in each space separately and determined their topological classifications. We showed that the addition of GB terms, dimensions, and other factors did not affect the topological classes of black holes in both spaces. The calculations and plots indicated that the AdS R-N and AdS EGB black holes showed similar behavior and their topological number sets in both spaces, i.e., BB and RPS, were similar and equal to ($W=+1$). However, AdS EPYM black holes show an interesting behavior. In addition to BBT and RPS, we also investigate the thermodynamic topology for this black hole in extended phase space thermodynamics (EPST). The changing ($r-\tau$) in both spaces "determined" similar behavior. Also, the total topological number for this black hole in the BBT, RPST and EPST "were" equal to ($W=0$). In fact, for the 4D EPYM black hole, the topological number and the total topological numbers are completely the same for all three spaces, i.e. $$\omega_{BBT}=\omega_{RPS}=\omega_{EPST}=+1, -1$$ or $$W_{BBT}=W_{RPS}=W_{EPST}=0$$\\\\
The present result may be due to the non-linear YM charge parameter and the difference between the gauge and gravity corrections in the above black holes. We know that Anindya Biswas obtained the EPYM AdS black hole solution in the 4D Einstein-Gauss-Bonnet (EGB) gravity, which is a higher-order modification of general relativity that includes a quadratic curvature term in the action. However, it is very interesting that these two structures show different behaviors in thermodynamic calculations in the BB and RPS, and that EPYM has a special behavior compared to EGB. Perhaps these behavioral discrepancies can be somehow traced to the gauge corrections. We know that the EPYM black hole theory is a solution in 4D anti-de Sitter (AdS) space with a nonlinear source, where the YM field is a generalization of the electromagnetic field that includes non-Abelian gauge symmetry and self-interactions. In fact, the main difference between EGB and EPYM is the source of the nonlinearity of the field equations. For EGB black holes, the nonlinearity comes from a higher-order curvature term, the Gauss-Bonnet term, which is a quadratic combination of the Ricci scalar, Ricci tensor, and Riemann tensor, and appears to be more of a correction. On the other hand, for EPYM black holes, the nonlinearity is due to the PYM term, which is the power of the Yang-Mills field strength tensor. Also, the GB term is a topological variable in four dimensions that does not contribute to the field equations unless coupled with a scalar field or a cosmological constant. But on the other hand, the PYM term is non-trivial in four dimensions and changes the field equations significantly. It would be very interesting to challenge the calculations related to thermodynamic topology from the perspective of the AdS/CFT correspondence by using CFT thermodynamics and comparing the results with other works. We could also pose a series of interesting questions for future works, such as:\\
- How does the thermodynamic topology of black holes in AdS space encode the phase structure and critical phenomena of the dual QFTs? What are the universal features and the model-dependent details of this correspondence?\\
- How can we extend thermodynamic topology to other physical systems, such as cosmological models, fluid dynamics, quantum information, etc.? What are the new topological structures and phenomena that emerge in these cases?
\section{Appendix A: Some points about $F$ method}
In this section, we will examine two very important points. In the first point, in order to check the effect of the dimensions and whether it affects the thermodynamic topology in the two BBT and RPS or not, we will study one of the previous black holes in six dimensions (6D EGB black hole). We will investigate the equations and plots in both spaces and analyze their results.
But for the second point, since the topological numbers ($\omega$) and total topological numbers ($W$) were completely similar in both spaces for the 4D EPYM black hole, therefore we decided to check the extended phase space thermodynamics (EPST) for this black hole and study its results with other spaces. In the end, we will explain the results of the work.
\subsection{6D EGB black hole in BBT}
Here, the entropy for 6D EGB black hole in BBT is as follows,
\begin{equation}\label{64}
S  = \frac{r_h^{4} \left(1+\frac{4 \alpha_{1}}{r_h^{2}}\right)}{4 G},l  = \frac{\sqrt{5}\, \sqrt{\frac{1}{P G \pi}}}{2}
\end{equation}
The temperature of the black hole in BBT is given by,
\begin{equation}\label{65}
T =\frac{\frac{5 r_h^{4}}{l^{2}}+3 r_h^{2}+\alpha_{1}-\frac{G \,q^{2}}{2 r_h^{4}}}{4 \left(r_h^{2}+2 \alpha_{1}\right) r_h \pi}
\end{equation}
The parameter G according to the mentioned equations, which is obtained as,
\begin{equation}\label{66}
G =\frac{2 r_h^{4} \left(3 r_h^{4}-3 r_h^{2} \alpha_{1}+2 \alpha_{1}^{2}\right)}{8 P \pi  r_h^{10}+48 P \pi  r_h^{8} \alpha_{1}+7 q^{2} r_h^{2}+10 q^{2} \alpha_{1}}
\end{equation}
We can obtain the mass and Helmholtz free energy for this black hole as follows,
\begin{equation}\label{67}
M =\frac{r_h^{3} \left(1+\frac{\alpha_{1}}{r_h^{2}}+\frac{r_h^{2}}{l^{2}}\right)}{4 G \pi}+\frac{q^{2}}{24 r_h^{3} \pi}
\end{equation}
and
\begin{equation}\label{68}
\mathcal{F }=\frac{r_h^{3} \left(1+\frac{\alpha_{1}}{r_h^{2}}+\frac{4 r_h^{2} P G \pi}{5}\right)}{4 G \pi}+\frac{q^{2}}{24 r_h^{3} \pi}-\frac{r_h^{4} \left(1+\frac{4 \alpha_{1}}{r_h^{2}}\right)}{4 G \tau}
\end{equation}
Therefore, ($\phi^{r_h}$) and $\phi^{\Theta}$ are obtained as,
\begin{equation}\label{69}
\begin{split}
&\phi^{r_h} = \frac{8 G P \pi  r_h^{8} \tau -8 \pi  r_h^{7}-16 \pi  r_h^{5} \alpha_{1}+6 r_h^{6} \tau +2 r_h^{4} \tau  \alpha_{1}-q^{2} G \tau}{8 G \pi  r_h^{4} \tau},
\\
&\phi^{\Theta}=-\frac{\cos \! \left(\Theta \right)}{\sin \! \left(\Theta \right)^{2}}
\end{split}
\end{equation}
Also, we will have,
\begin{equation}\label{70}
\tau =\frac{8 r_h^{5} \pi  \left(r_h^{2}+2 \alpha_{1}\right)}{8 G P \pi  r_h^{8}+6 r^{6}+2 r_h^{4} \alpha_{1}-G \,q^{2}}
\end{equation}

\begin{figure}[h!]
 \begin{center}
 \subfigure[]{
 \includegraphics[height=5.5cm,width=6cm]{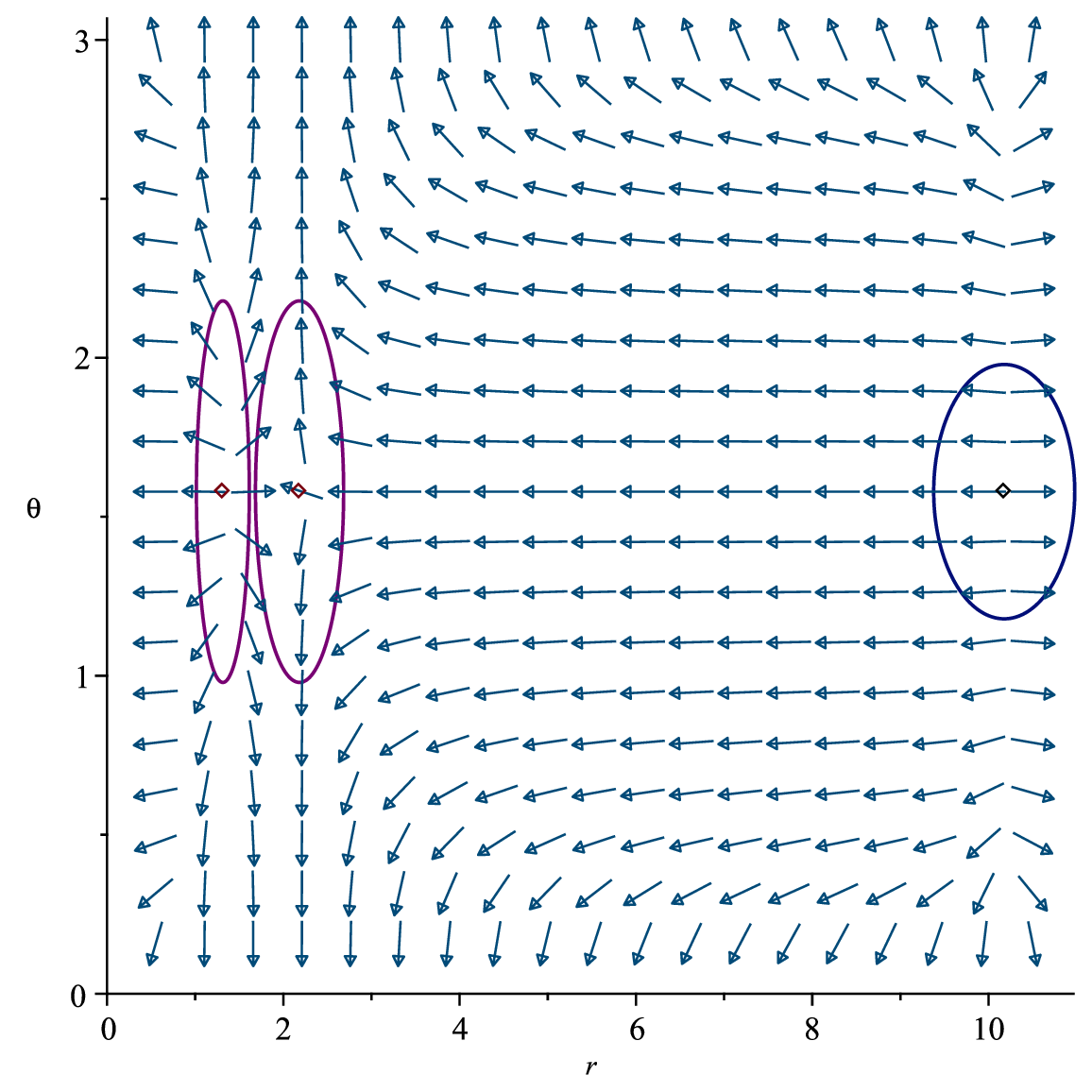}
 \label{7a}}
 \subfigure[]{
 \includegraphics[height=5.5cm,width=6cm]{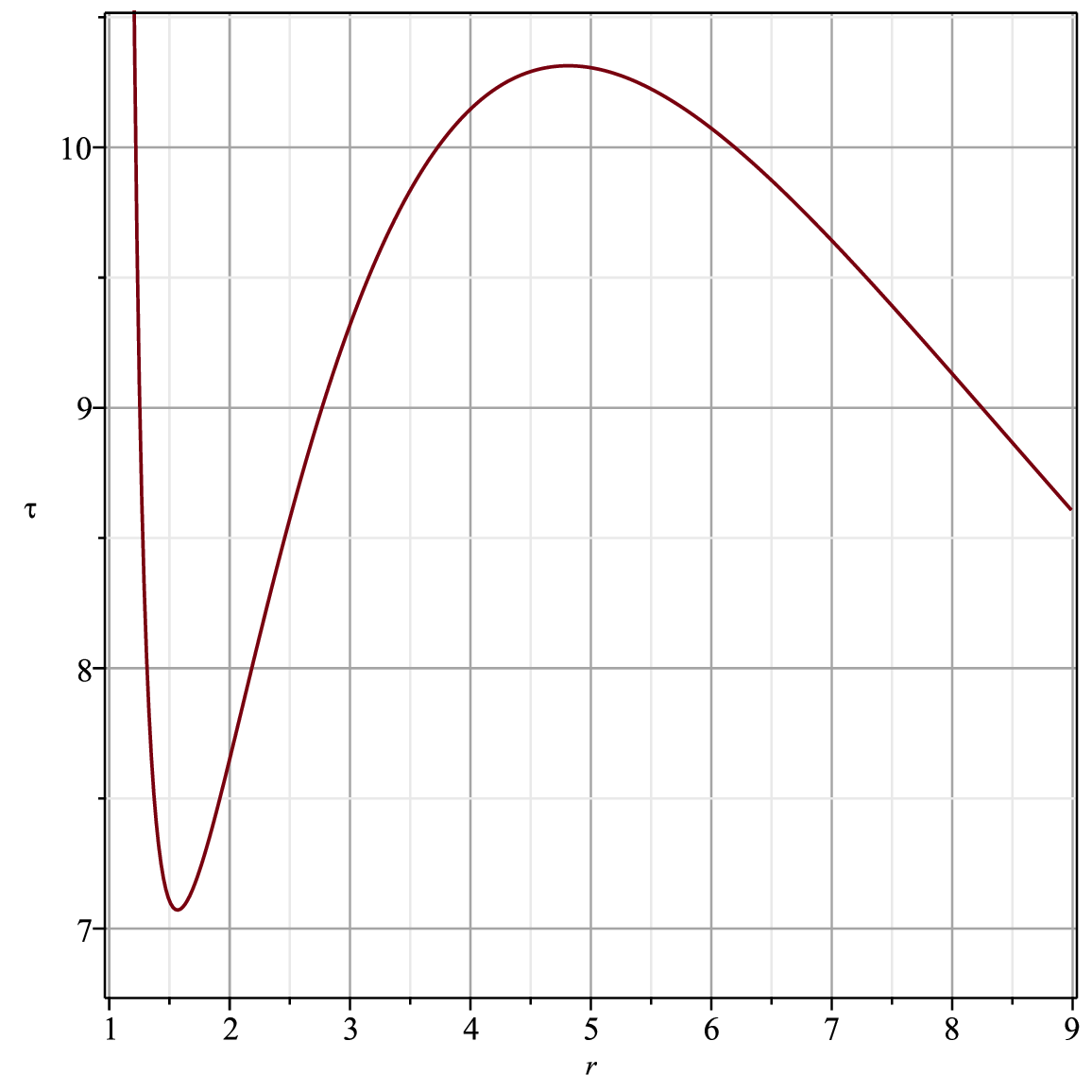}
 \label{7b}}
  \caption{\small{
The vector field $n$ on a part of the $(r-\Theta)$ plane for the 6D EGB black hole in TBHT with $(\hat{q}=10, p=0.1, G=0.1, \alpha=0.1)$ is shown by the blue arrows in Fig (7a). The ZPs are at $(r,\Theta)=(1.313885737, 1.57), (2.183792565, 1.57)$ and $(10.17775329, 1.57)$. We choose three closed loops (purple loop) and (blue loop) that encircle the ZPs. Fig (7b) shows the curve of equation (70).}}
 \label{7}
 \end{center}
 \end{figure}

\subsection{6D EGB black hole in RPS}
Also, the entropy and tempruture of 6D EGB black hole in RPS is as follows,
\begin{equation}\label{71}
S =\frac{r_h^{4} \left(1+\frac{4 \alpha_{1}}{r_h^{2}}\right)}{4 G},l  = \frac{\sqrt{5}\, \sqrt{\frac{1}{P G \pi}}}{2},q_{1} = \sqrt{C}\, q
\end{equation}
and
\begin{equation}\label{72}
T =\frac{\frac{5 r_h^{4}}{l^{2}}+3 r_h^{2}+\alpha_{1}-\frac{G \,q^{2}}{2 r_h^{4}}}{4 \left(r_h^{2}+2 \alpha_{1}\right) r_h \pi}
\end{equation}
The parameter C with respect to above mentioned equations, which is given by,
\begin{equation}\label{73}
C =\frac{\sqrt{2}\, \sqrt{\left(3 l^{2} r_h^{4}-5 r_h^{6}-3 l^{2} r_h^{2} \alpha_{1}-30 \alpha_{1} r_h^{4}+2 l^{2} \alpha_{1}^{2}\right) \left(7 r_h^{2}+10 \alpha_{1}\right)}\, q_{1} l^{3}}{2 \left(3 l^{2} r_h^{4}-5 r_h^{6}-3 l^{2} r_h^{2} \alpha_{1}-30 \alpha_{1} r_h^{4}+2 l^{2} \alpha_{1}^{2}\right) r_h^{2}}
\end{equation}
The mass and Helmholtz free energy for this black hole which are calculated,
\begin{equation}\label{74}
M =\frac{\left(\left(6 r_h^{6}+6 \alpha_{1} r_h^{4}\right) l^{2}+6 r_h^{8}\right) C^{2}+q_{1}^{2} l^{6}}{24 l^{6} \pi  C \,r_h^{3}}
\end{equation}
and
\begin{equation}\label{75}
F =\frac{\left(\left(6 r_h^{6}+6 \alpha_{1} r_h^{4}\right) l^{2}+6 r_h^{8}\right) C^{2}+q_{1}^{2} l^{6}}{24 l^{6} \pi  C \,r_h^{3}}-\frac{r_h^{4} C \left(1+\frac{4 \alpha_{1}}{r_h^{2}}\right)}{4 l^{4} \tau}
\end{equation}
Therefore, $(\phi^{r_h})$ and $\phi^{\Theta}$ are obtained as,
\begin{equation}\label{76}
\begin{split}
&\phi^{r_h} = \frac{-8 \left(\left(r_h^{3} \pi -\frac{3}{4} r_h^{2} \tau +2 r_h \pi  \alpha_{1}-\frac{1}{4} \tau  \alpha_{1}\right) l^{2}-\frac{5 r_h^{4} \tau}{4}\right) r_h^{4} C^{2}-l^{6} \tau  q_{1}^{2}}{8 l^{6} \pi  r_h^{4} C \tau},
\\
&\phi^{\Theta}=-\frac{\cos \! \left(\Theta \right)}{\sin \! \left(\Theta \right)^{2}}
\end{split}
\end{equation}
So, $\tau$ for 6D EGB black hole in RPS is obtained as follows,
\begin{equation}\label{77}
\tau =\frac{8 C^{2} \pi  l^{2} r_h^{5} \left(r_h^{2}+2 \alpha_{1}\right)}{6 C^{2} l^{2} r_h^{6}+10 C^{2} r_h^{8}+2 C^{2} l^{2} r_h^{4} \alpha_{1}-q_{1}^{2} l^{6}}
\end{equation}

\begin{figure}[h!]
 \begin{center}
 \subfigure[]{
 \includegraphics[height=5.5cm,width=6cm]{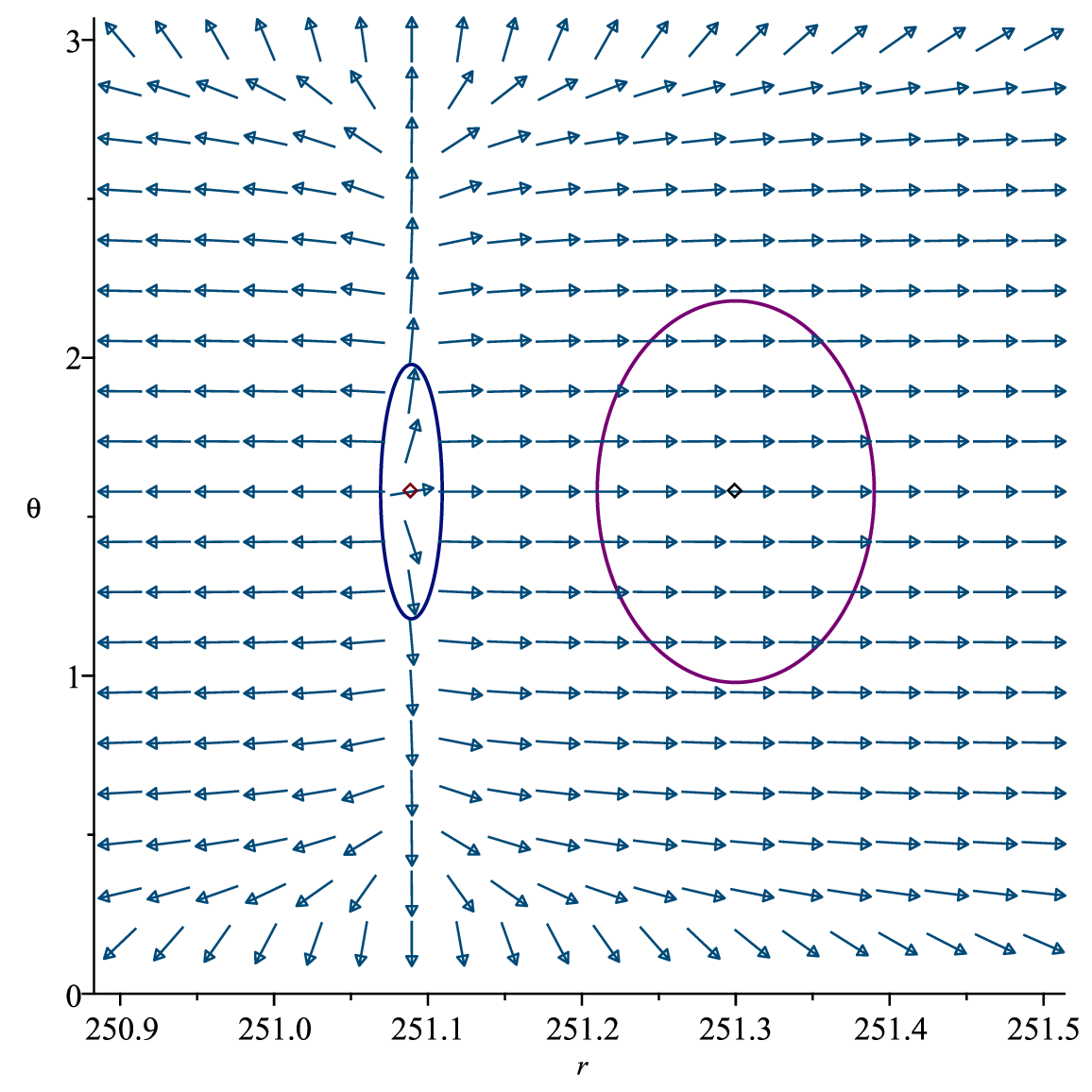}
 \label{7a}}
 \subfigure[]{
 \includegraphics[height=5.5cm,width=6cm]{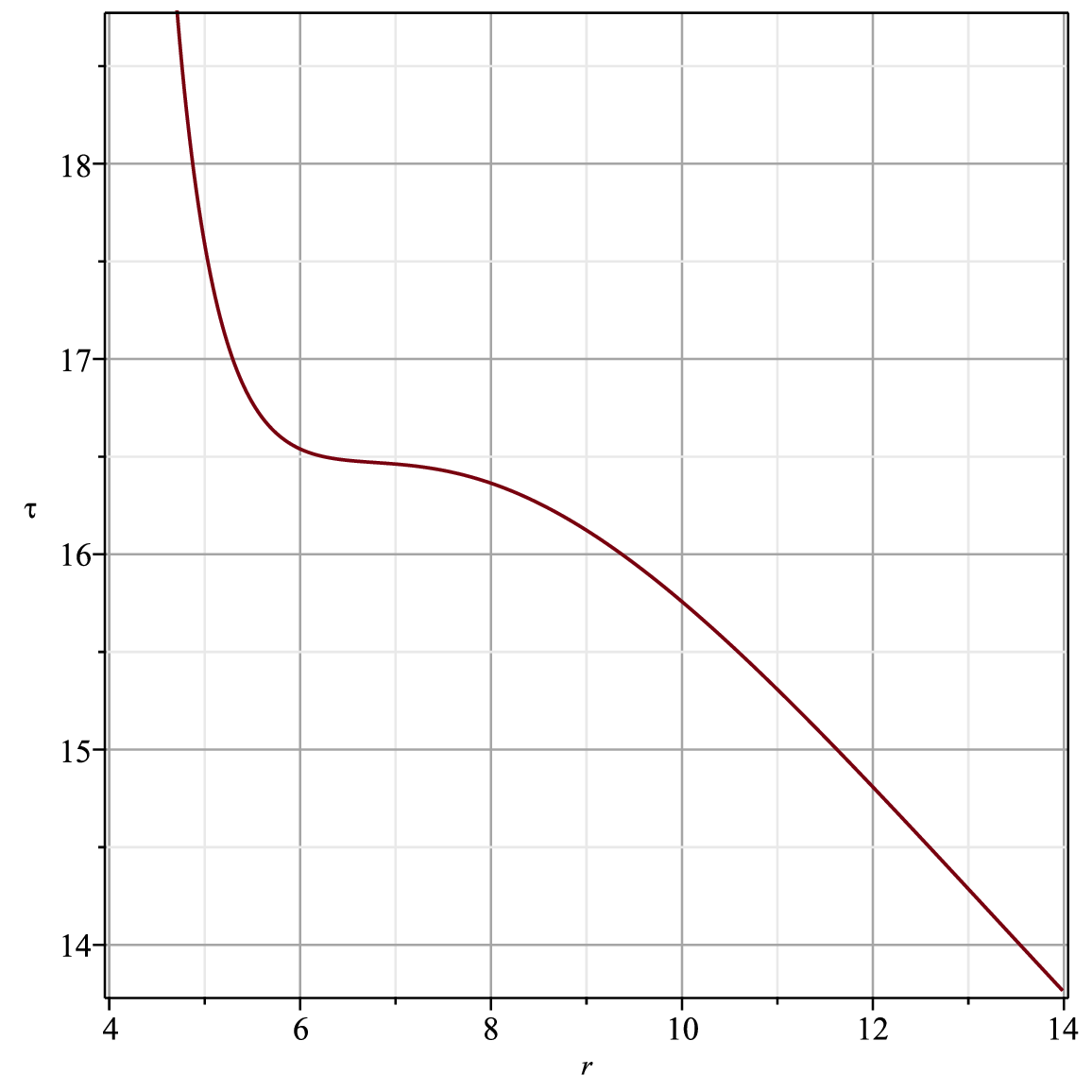}
 \label{7b}}
  \caption{\small{
The vector field $n$ on a part of the $(r-\Theta)$ plane for the 6D EGB black hole in RPST with $(\hat{q}=10, l=10, C=7.3, \alpha=0.1)$ is shown by the blue arrows in Fig (7a). The ZP is at $(r,\Theta)=(251.0892506, 1.57)$. We choose two closed loops $Z_1$ (purple loop) and $Z_2$ (blue loop), where $Z_2$ encircles the ZP but $Z_1$ does not. Fig (7b) shows the curve of equation (77).}}
 \label{7}
 \end{center}
 \end{figure}

\subsection{4D EPYM black hole in EPST}
Here, we will rewrite the equations for 4D AdS EPYM black holes in EPST. So with respect to\cite{}, the parameters $T$, $P$ and $M$ are calculated
\begin{equation}\label{78}
T =\frac{1+8 r_h^{2} P \pi -\frac{\left(2 q^{2}\right)^{\gamma}}{2 r_h^{4 \gamma -2}}}{4 r_h \pi}
\end{equation}
and
\begin{equation}\label{79}
P =\frac{-8 \,2^{-1+\gamma} \left(q^{2}\right)^{\gamma} r_h^{-4 \gamma +2} \gamma +2^{\gamma} \left(q^{2}\right)^{\gamma} r_h^{-4 \gamma +2}+2}{16 r_h^{2} \pi}
\end{equation}
The mass is as fokkows,
\begin{equation}\label{80}
M =\frac{r_h \left(1+\frac{r_h^{2}}{l^{2}}-\frac{q^{2 \gamma} 2^{-1+\gamma}}{r_h^{4 \gamma -2} \left(4 \gamma -3\right)}\right)}{2}
\end{equation}
Also, the entropy and AdS radius which is given by,
\begin{equation}\label{81}
\begin{split}
&S = r_h^{2} \pi ,\\
&l  = \frac{\sqrt{6}\, \sqrt{\frac{1}{\pi  P}}}{4}
\end{split}
\end{equation}
Therefore, the Helmholtz free energy in EPST is given by,
\begin{equation}\label{82}
\mathcal{F} =\frac{r_h \left(1+\frac{8 r_h^{2} P \pi}{3}-\frac{q^{2 \gamma} 2^{-1+\gamma}}{r_h^{4 \gamma -2} \left(4 \gamma -3\right)}\right)}{2}-\frac{r_h^{2} \pi}{\tau}
\end{equation}
We compute the vector fields $\phi^{r_h}$ and $\phi^{\Theta}$ as bellow,
\begin{equation}\label{83}
\begin{split}
&\phi^{r_h}=\frac{16 r_h^{2} P \pi  \tau +2^{\gamma} q^{2 \gamma} r_h^{-4 \gamma +2} \tau -8 r_h \pi +2 \tau}{4 \tau},\\
&\phi^{\Theta} = -\frac{\cos \! \left(\Theta \right)}{\sin \! \left(\Theta \right)^{2}}
\end{split}
\end{equation}
Also, we will have,
\begin{equation}\label{84}
\tau =\frac{8 r_h \pi}{16 r_h^{2} P \pi +2^{\gamma} q^{2 \gamma} r_h^{-4 \gamma +2}+2}
\end{equation}

\begin{figure}[h!]
 \begin{center}
 \subfigure[]{
 \includegraphics[height=5.5cm,width=6cm]{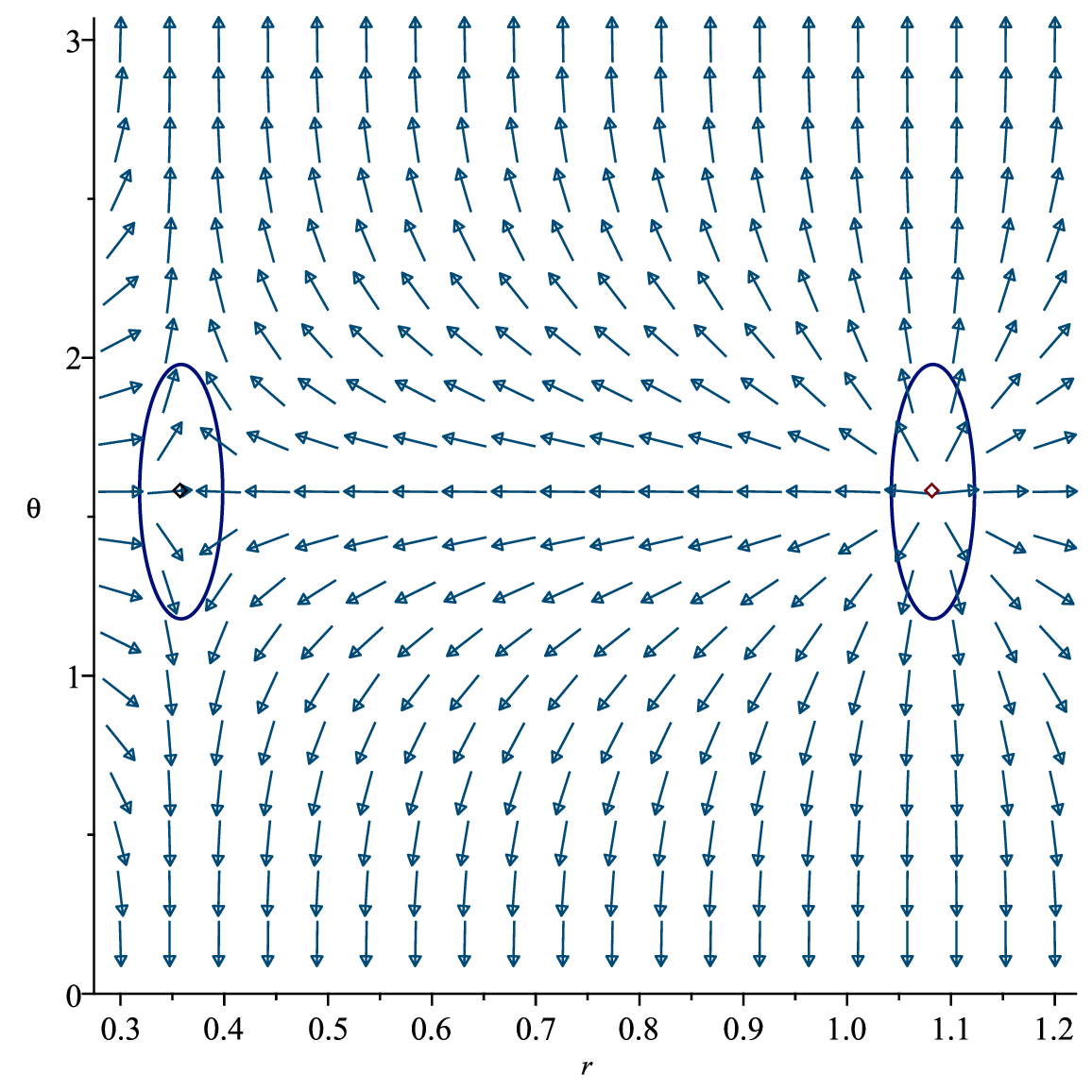}
 \label{8a}}
 \subfigure[]{
 \includegraphics[height=5.5cm,width=6cm]{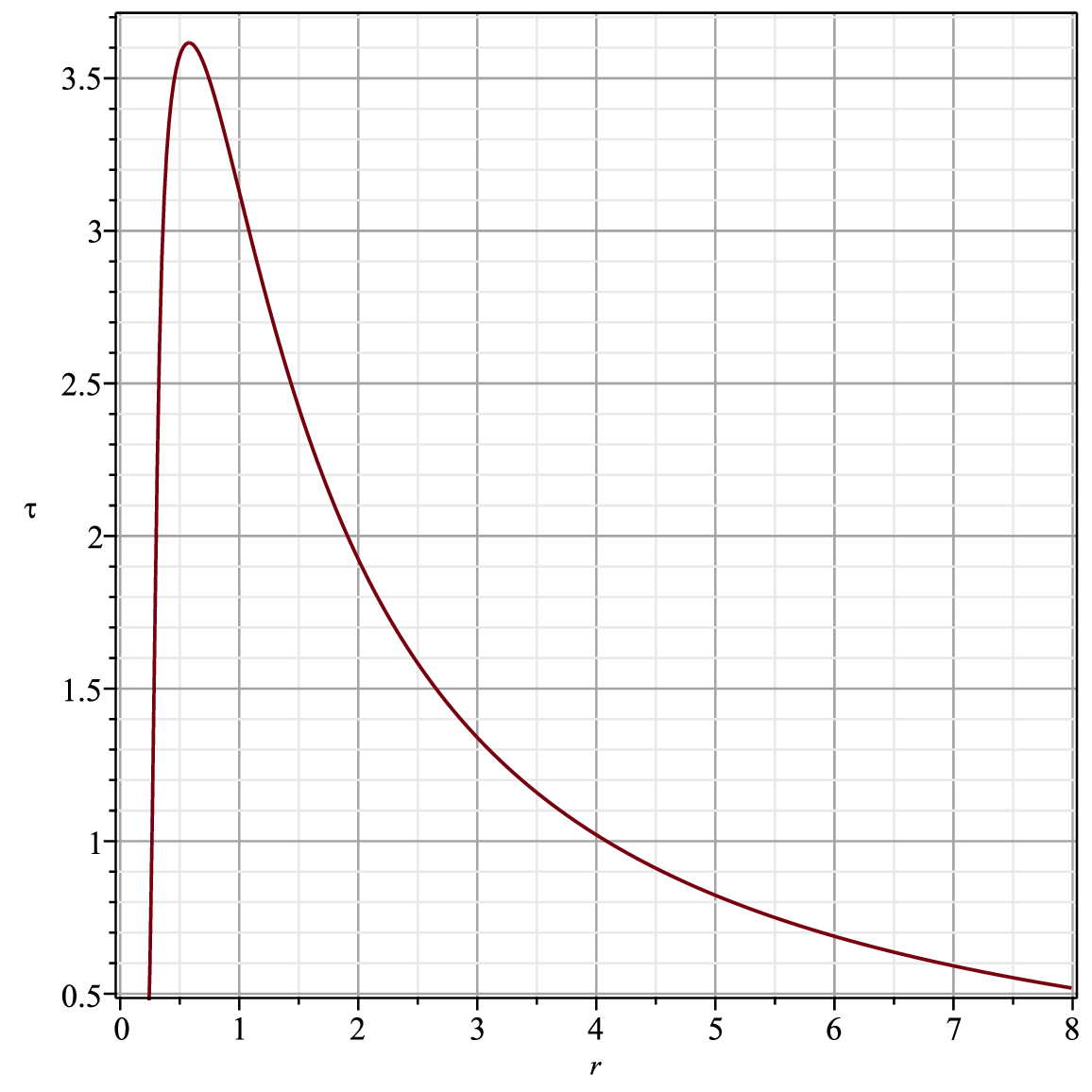}
 \label{8b}}
  \caption{\small{
The vector field $n$ on a part of the $(r-\Theta)$ plane for the AdS EPYM black holes with $(\hat{q}=0.1, p=0.12,  \gamma=3)$ is shown by the blue arrows in Fig (8a). The ZPs are at $(r,\Theta)=(0.3585003619, 1.57), (1.082619261, 1.57)$. We choose two closed loops (blue loop) that encircle the ZPs. Fig (8b) shows the curve of equation (84).}}
 \label{8}
 \end{center}
 \end{figure}
As can be seen in all three plots 7, 8, and 9, they show similar behaviors to their previous cases.
Figure (7a) shows the three topological numbers ($\omega+1, -1, +1$) with the total topological numbers ($W=1$) according to the free parameters mentioned for (6D EGB black hole in BBT), which is similar to figure (2a) in 4 dimensions. Also, figure (8a) shows only one positive topological number +1 for the mentioned black hole in RPS, which is completely similar to figure (5a).
Hence, it can be concluded that the dimension has no effect on the topological number and thermodynamic topology.
But the valuable point is related to (the 4D EPYM black hole). As it is clear, figure (9a) determines the topological numbers ($\omega= -1, +1$) with the total topological charge ($W=0$) for this black hole in EPST, which has a completely similar behavior to figures (3a) and (6a).
In fact, for the 4D EPYM black hole, the topological number and the total topological numbers are completely the same for all three spaces, which is a very valuable point, i.e. $(\omega_{BBT}=\omega_{RPS}=\omega_{EPST}=+1, -1)$ or $W_{BBT}=W_{RPS}=W_{EPST}=0$. Figures (7b), (8b), and (9b) have the same behavior as their previous cases.

\end{document}